\documentclass{ws-ijmpd}

\begin{document}

\markboth{E. Kafexhiu, F. Aharonian and G. Vila}
{Nuclear reactions in hot astrophysical plasmas}

\catchline{}{}{}{}{}
%
%%%%%%%%%%%%%%%%%%%%%%%%%%%%%%%%%%%%%%%%%%%%%%%%%%%%%%%%%%%%%%%%%%%%

% \title{INSTRUCTIONS FOR TYPESETTING 
% MANUSCRIPTS\footnote{For the title, try not to use more than 3 lines. 
% Typeset the title in 10~pt Times roman, uppercase and boldface.}  }

\title{Nuclear reactions in hot astrophysical plasmas with $T>10^{10}$ K}

\author{ERVIN KAFEXHIU\footnote{Fellow of the International Max Planck Research School for Astronomy and Cosmic Physics at the University of Heidelberg (IMPRS-HD)}}

\address{Max-Planck-Institut f\"ur Kernphysik\\
Saupfercheckweg 1, Heidelberg, 69117, Germany\\
ervin.kafexhiu@mpi-hd.mpg.de}

\author{FELIX AHARONIAN}

\address{Max-Planck-Institut f\"ur Kernphysik\\ 
Saupfercheckweg 1, Heidelberg, 69117, Germany\\
Dublin Institute for Advanced Studies, 31 Fitzwilliam Place, Dublin 2, Ireland\\
felix.aharonian@mpi-hd.mpg.de}

\author{GABRIELA S. VILA}

\address{Instituto Argentino de Radioastronom\'ia (IAR - CONICET)\\
C.C. N$^\circ$ 5 (1894), Villa Elisa, Buenos Aires, Argentina\\
gvila@iar-conicet.gov.ar}
\maketitle

\begin{history}
\received{Day Month Year}
\revised{Day Month Year}
\comby{Managing Editor}
\end{history}

\begin{abstract}
The importance of nuclear reactions in low-density astrophysical plasmas with ion temperatures $T \geq 10^{10}$ K has been recognized for more than thirty years. However, the lack of  comprehensive data banks of relevant nuclear reactions and the limited computational power have not previously allowed detailed theoretical studies. 
Recent developments in these areas make it timely to conduct comprehensive studies on the nuclear properties of very hot plasmas formed around compact relativistic objects such as black holes and neutron stars. Such studies are of great interest in the context of scientific programs of future low-energy cosmic $\gamma$-ray spectrometry.
In this work, using the publicly available code TALYS, we have built a large nuclear network relevant for temperatures exceeding $10^{10}$ K. We have studied the evolution of the chemical composition and accompanying prompt gamma-ray emission of such high temperature plasmas.
We present the results on the abundances of light elements D, T, $^3$He, $^4$He, $^{6}$Li, $^{7}$Li $^{9}$Be, $^{10}$B,  $^{11}$B, and briefly discuss their implications on the astrophysical abundances of these elements.
\end{abstract}
% We then study the chemical and prompt $\gamma$-ray lines emission and spectra evolution of these high temperatures plasmas due to the nuclear reactions.suggest some mechanisms which could produce higher abundances of these light elements in astrophysical scenarios.

\keywords{Nuclear reactions, nucleosynthesis, Line: profiles, Gamma rays: general}

%\keywords{Nuclear reactions, nucleosynthesis, abundances, Line: formation, Line: identification,
%Line: profiles, Plasmas, Radiation mechanisms: general, Shock waves, Accretion, Gamma rays: general}

\section{Introduction} \label{sec:intro}
Low-density optically thin, hot  plasmas with ion temperature exceeding 1 MeV can form near compact relativistic objects, such as in accretion flows close to black holes\cite{shap} and in strong shock waves related, for example, to supernova explosions.\cite{col} The most straightforward approach to studying the dynamics of  formation and evolution of such plasmas is the detection of prompt $\gamma$-ray lines which are unambiguous signatures of specific nuclei.

Collisions in thermal plasmas with temperatures exceeding 1~MeV are characterized not only by excitation reactions with emission of gamma-ray lines, but also by the destruction of nuclei. Generally, in such plasmas, the destruction of nuclei proceeds on timescales shorter than excitations ones. Thus, only prompt de-excitation lines related to specific nuclei are expected to contribute to the plasma radiation.\cite{AS84} The presence of a given prompt $\gamma$-ray line is related to the existence of a given nucleus, which is itself governed by the destruction processes. This suggests that there should exist a strong relation between the gamma-ray line spectrum and nuclear destruction processes.

The break-up processes are sensitive to the plasma temperature and density. Therefore, the detection of $\gamma$-ray lines and their spectral profile evolution may reveal unique information about the chemical composition and physical parameters of the plasma, such as temperature and density at any given time.

The first attempts to assess the importance of nuclear processes in hot, thin astrophysical plasmas were made in Refs.~\refcite{AS84},~\refcite{AS87}. Different astrophysical implications of nuclear reactions in hot two-temperature plasmas have been discussed also in Ref. ~\refcite{RamaRees85} - \refcite{Spruit10}. These works provide estimates for the isotope ratios of different light elements such as Deuterium (D),  $^{6}$Li, $^{7}$Li $^{9}$Be, $^{11}$B, etc, and also estimates for the $\gamma$-ray line luminosity due to nuclear reactions. A mechanism to produce light elements, alternative to Big-Bang Nucleosynthesis, is also suggested in Refs.~\refcite{AS84},~\refcite{AS87}. In particular, they discuss a mechanism for deuterium production in hot plasmas. 

% There, the contribution of strong gamma-ray lines to the luminosity of accretion disks and spherically symmetric accretion flows is estimated. According to the results of those works, even under the most favorable conditions of temperature and composition, the gamma-ray line luminosity would be low, and undetectable with the instruments available in the epoch. 

In these early works, however, only a limited number of reactions could be taken into account, mainly due to the lack of data on nuclear cross sections and the increased computational complexity of calculations as more nuclei are considered.

Most of these difficulties have nowadays been overcome. The introduction of on-line nuclear databases\cite{ENDF}$^,$\cite{EXFOR} provides easy access to experimental and theoretical data on nuclear cross sections. There also exist publicly available computer codes that allow the calculation of nuclear cross sections and other quantities of interest. These make use of several theoretical models.

In this work we build a large nuclear reaction network and use it to calculate the temporal evolution of the composition, the $\gamma$-ray line emissivity, and the corresponding $\gamma$-ray spectrum, for a low density plasma with ion temperature $T\geq 1$~MeV. Two cases are considered: when the temperature and density evolve with time, and when they remain constant. We present results for the abundance of light elements and the baryonic $\gamma$-ray spectrum due to nuclear reactions and neutron-proton Bremsstrahlung. Finally, we discuss some mechanisms that could enhance the production of light elements. This could be of special importance, because it could modify the cosmic abundance of light elements predicted by the Standard Big-Bang Nucleosynthesis model. The production of neutrons is also of considerable interest, since these particles can escape from the source and interact with nearby targets.

\section{Cross sections and reaction rates}
\label{sec:cross}

The calculation of the evolution of the nuclear abundances and the $\gamma$-ray line emission in low density, high temperature plasmas, requires the use of a comprehensive nuclear network. This network should take into account all relevant nuclear reactions that may occur in a hot plasma with $T\geq 1$~MeV. 

There is an extensive literature about how to build and solve nuclear networks, e.g. see  Refs.~\refcite{FCZ} - \refcite{hixtil}. Nowadays, all this knowledge is compressed and very easy to access through on-line nuclear databases for astrophysical applications, such as {\it JINA Reaclib},\cite{jina} {\it NON-SMOKER},\cite{nonsmok} {\it NucAstroData},\cite{nucastro} etc. There also exist many computer programs which store, manage and solve the networks, e.g. {\it libnuceq} and {\it libnucnet}.\cite{clemson} 
However, all these nuclear networks are designed to be applied in scenarios such as big-bang nucleosynthesis, core collapse supernova, stellar interiors, etc. Generally, they are valid for temperatures $T\le10^{10}$~K. Therefore, to describe nuclear reactions for astrophysical plasmas with $T\ge10^{10}$~K (or $T\ge1$~MeV), we find it necessary to extend the aforementioned works.
% 
% In principle, the network should include all possible interactions between the nuclear species initially present or those potentially created in the plasma, and all their resulting channels. In practice not all the channels are important. Many channels are rare or less probable, compare to other channels which are the most dominant and which will play the major role in the plasma nuclear evolution. A wise selection of reactions and their resulting channels, will pay off immediately in the computing time and the solution accuracy. In this first attempt, we have parametrized the channel selection procedure -- as it will be discussed later on -- by putting a threshold in the channel production reaction-rate.

Without loosing generality, our main purpose is to describe the evolution of the abundance of nuclear species and the $\gamma$-ray emission of a very hot, low density and optically thin astrophysical plasma. To further simplify the problem, we assume the following conditions hold:

\begin{itemize}
\item all nuclei are instantaneously thermalized, and their velocity distributions are described by a Maxwellian distribution function at temperature $T$,
\item all target nuclei interact in their ground state,
\item the plasma has low density ($n\sim10^{18}\,\rm{cm^{-3}}$) and high temperature ($T\ge1$~MeV), and its initial composition is similar to the standard cosmic/solar chemical abundances.
\end{itemize}

These assumptions dramatically simplify the reaction network. The first assumption reduces the calculation of the reaction rates to a simply Maxwellian average $<\sigma\,v>$. Thus, $<\sigma\,v>$ will be only a function of the temperature $T$ as shown in Ref.~\refcite{FCZ}:
\begin{equation}\label{rf}
<\sigma\,v>_{ij}^k(T)=\sqrt{\frac{8}{\pi\, \mu_{ij} (k_BT)^3}}\,\int_0^\infty  \,\sigma_{ij}^k(E)\,E\, \exp\left(-\frac{E}{k_{\rm{B}}T}\right)\,dE,
\end{equation}
where $E$ is the projectile CM-frame energy, $\sigma_{ij}^k$ is the cross section of the binary interaction $i+j\to k+...$, $\mu_{ij}$ is the reduced mass of the interacting system, and $k_{\rm{B}}$ is the Boltzmann constant.

The second assumption simplifies both the calculation of the cross sections and the reaction rates. In a high temperature plasma, different isomeric states of a given nucleus may be in thermal equilibrium. Therefore, to obtain the reaction rate, one should sum the contributions of all isomers.\cite{WFHZ}$^,$\cite{RauThil}$^-$\cite{rauscher}  As a first approach, we have assumed here that all isomeric states decay before the nuclei interact. 

The third assumption restricts the number of targets/channels and interactions to be taken into account.

Under these assumptions, and since we are interested in low density plasmas, only the binary interactions contribute to the chain of nuclear reactions. For the composition of plasmas similar to standard astrophysical environments, only the reactions with participation of protons ($p$), neutrons ($n$), and alpha particles ($\alpha$) are important, c.f. \ref{app:assum}.

% we have estimated that the most important nuclear reactions are the binary ones, where at least one of the particles is a {\it proton} ($p$), a {\it neutron} ($n$), or an {\it alpha} ($\alpha$) particle, c.f. the \ref{app:assum}. 

Although the nuclear reaction rates for $p$, $n$ and $\alpha$ projectiles can be simply calculated from the local conditions of the plasma - such as ion temperature and density - this is not the case for the interactions of photons with nuclei. The $\gamma$-interaction reaction rates not only depend on local parameters, but also on the number density of photons, which itself depends on the optical depth of the plasma. Furthermore, the optical depth is also a function of ``non-local'' variables such as the electron number density and size of the plasma. 

The $\gamma$-rays are produced and transported inside the plasma. Hot two-temperature plasmas radiate X-rays and $\gamma$-rays through interactions of both components of the plasma. The dominant radiative cooling channels for ions are the excitation of nuclei with production of prompt $\gamma$-ray lines, the $\gamma$-radiation released at $p-n$ capture, as well as $p-n$ Bremsstrahlung. The electronic component radiates more efficiently, predominantly through Bremsstrahlung, and in the case of large optical depth, is cooled due to comptonization.

% The major radiative cooling of electrons is due to electron bremsstrahlung Photons coming from electrons however, are dominated by {\it e-e} or {\it e-ion} bremsstrahlung. One must note, that for plasmas with electron temperature comparable with ion temperature, $\gamma$-radiation coming from electrons dominates. If electron temperature is much lower than ion temperature, the most dominant photon production process will be {\it p-n} capture and bremsstrahlung, and the photon production rate in this case will be 2 to 3 orders of magnitude lower then the case of plasma with electron temperature comparable to the ion temperature.

Because the photon number density depends on the optical depth of the plasma, the effect of the photodisintegration of nuclei cannot be estimated without considering the plasma as a whole. The importance of this effect must be checked for each specific problem. In this paper we have neglected photodisintegration. Generally, this is a good approximation for optically thin, low density plasmas. We illustrate in \ref{app:gamma} the effect of the photodisintegration on $^{16}$O and Deuterium nuclei, for the case of a spherically symmetric and optically thin plasma. We show that even if the electron temperature is comparable with the ion temperature, the effect of photodisintegration on the evolution of the plasma can still be safely neglected. However, this might not be the case for an optically thick plasma, or for the case where nuclear statistical equilibrium is reached. 

The integral in Eq.(\ref{rf}) is evaluated over the energy range ($0$, $\infty$). In practice, however, one needs to integrate Eq.(\ref{rf}) only in a ``narrow'' energy interval from which comes the most important contribution to the reaction rate. To have an accurate estimation of the reaction rate we must know precisely the cross sections in that narrow energy range, since the contribution of the leftover interval is negligible, by definition. As discussed in \ref{app:ener}, in the temperature range 0.1-20~MeV the most relevant interval of projectile energy is that between 0.01 MeV and 150 MeV.

The Maxwellian average $<\sigma\,v>$ is the main ingredient to solve the kinetic equations.\cite{WFHZ} It is convenient though, to work with mass fraction abundances instead of number densities. If we define the mass fraction abundance of the element $i$ as $X^i=A_i\,n^i/\rho$, where  $A_i$ is the mass number, and $\rho\equiv\sum_l\,A_l\,n^l$ the nucleon density,\footnote{In practice, nucleon density $\rho$ can be converted to mass density $\rho_{mass}$, simply by multiplying it with nucleon mass $m_u$, $\rho_{mass}\approx\rho\cdot m_u$} then for our case the kinetic equation reads:
\begin{equation}\label{eq:abund}
\dot{X}^k\,=\,\sum_i^{n,p,\alpha}\,\sum_j^N\,\frac{\rho}{1+\delta_{ij}}\left(\frac{A_k}{A_i\,A_j}\right) \,<\sigma\,v>_{ij}^k\,X^i\,X^j.
\end{equation}
The index $i$ runs over the three types of projectiles (protons, neutrons and alpha particles), whereas $j$ runs over all $N$ elements of the network. 

The remaining step for calculating $<\sigma\,v>_{ij}^k$ in Eq.(\ref{rf}), is to provide the cross sections in the relevant projectile energy interval. We chose to use the public available code TALYS\cite{talys} to calculate the cross sections $\sigma_{ij}^k$ for all relevant nuclear interactions, together with the $\gamma$-ray lines associated with these interactions. TALYS is an attractive tool to be applied to the study of nuclear processes in high-temperature plasmas because:
\begin{description}
\item[i.] the energy of the projectile can be in the range 1 keV-200 MeV,
\item[ii.] it incorporates modern nuclear models for the optical model, level densities, direct reactions, compound reactions, pre-equilibrium reactions,
\item[iii.] it can treat $\gamma$, $n$, $p$, D, T, $^3$He, and $\alpha$-particles as both projectiles and ejectiles,
\item[vi.] target mass numbers can be between $A = 12$ and $A = 339$. The code can also calculate cross section for targets with mass number in the range $A = 6-12$, but the results may not be as accurate.
 \end{description}
% For masses with $A<12$ another source of cross sections (or rates) is needed.

We used TALYS to calculate the cross sections for reactions involving elements from $^6$Li to $^{70}$Zn. For targets with \mbox{ $A\leq$ 6} (i.e. $n$, $p$, D, T, $^3$He, $^4$He), we collected the cross sections from theoretical evaluations; when no evaluations were available we used experimental data. Both evaluations and experimental data points were taken from the publicly available databases ENDF\cite{ENDF} and EXFOR.\cite{EXFOR} 

The reaction network should in principle satisfy a ``closure condition'': the reaction products of all nuclei in the network must be also part of the same network. In practice we select the most relevant reactions/channels, which implies that the closure condition holds only approximately. We can fix the accuracy up to which the closure condition is satisfied by introducing a threshold value for the Maxwellian average. Every channel with a Maxwellian average smaller than the threshold is ignored. Changing the value of the threshold changes the number of reaction channels in consideration, therefore this parameter plays a crucial role in the accuracy of the solution, the calculation time, the dimension of the network, etc. 

To calculate the cross sections and reaction rates for nuclei from $^6$Li to $^{70}$Zn we proceeded as follows. Using TALYS we calculated the cross sections for the interaction of solar abundance nuclei (the stable nuclei) from $^6$Li to $^{70}$Zn, with $n$, $p$ and $\alpha$ as projectiles. Then, the corresponding reaction rates were calculated as in Eq.(\ref{rf}). At $T=1$~MeV the largest reaction rates are of the order of $R_{ij}^k\sim 10^{-16}-10^{-15}\;\rm{cm}^3\,\rm{s}^{-1}$. We chose a threshold value $R^{\rm{th}}= 10^{-19}\;\rm{cm}^3\,\rm{s}^{-1}$, ensuring that the contribution of the dismissed reactions would be no more than $0.1\%$.

From the selected channels new targets may appear. They are potentially created in the plasma due to the reactions of the previous generation of nuclei, and enter in the calculations as a second target generation. We then repeated the previous procedure for the new targets, i.e. calculating the cross sections, the reaction rates, and the selection of the relevant channels according to the chosen threshold value. The new channels may still introduce a third generation of new nuclei. We applied the algorithm until there is no generation of new nuclear species. The network is then closed up to the accuracy given by $R^{\rm{th}}$. Notice that the value of $R^{\rm{th}}$ must be adjusted according to the scope of the problem, and there is no general advice for its choice.

We extended our calculations by fitting the Maxwellian averages $<\sigma\,v>_{ij}^k$ as a function of temperature using the parameterization with seven parameters suggested in Ref.~\refcite{RauThil}, Ref.~\refcite{ThTrAr} : 
\begin{equation}\label{fits}
<\sigma\,v>_{ij}^k= \exp(a_1 + a_2\,T^{-1} + a_3\,T^{-\frac{1}{3}} + a_4\,T^{\frac{1}{3}} + a_5\,T + a_6\,T^\frac{5}{3} + a_7\,\log T).
\end{equation}
\vspace{0.2cm}

\noindent The fit is valid for $T\in[0.1,\,20]$ MeV.  We estimated the goodness of the fit for the seven parameters using the $\chi^2$ method. In the mentioned temperature range the majority of the rates have a residual fluctuation less than 5\%.

As we have already mentioned, for light elements with $A\le6$  we obtained the relevant cross sections from evaluations or, in the case where no evaluations were found, from experimental data points. It must be mentioned here that using evaluations or experimental data introduces additional uncertainties. Some evaluations do not exceed projectile energies of 20 MeV. In these cases, we extrapolated the cross section in the high energy region as $\sigma\sim 1/E$, see e.g. Ref.~\refcite{RaKoLi}. Listed below are all the reactions involving nuclei with $A\le6$ we have included in this work:\\
 
\begin{tabular}{lllll}
H1) $^1$H (n, $\gamma$) D & &  & & \\
 \\

D1) D (n, 2n) $^1$H & &
D3) D (p, np) $^1$H & &
D5) D ($\alpha$, np) $^4$He \\
D2) D (n, $\gamma$) T  & & 
D4) D (p, $\gamma$) $^3$He & &
% d6) D ($\alpha$, $\gamma$) $^6$Li \\
\\
\\

T1) T (n, 2n) D & &
T3) T (p, n) $^3$He & &
T5) T (p, $\gamma $) $^4$He \\
T2) T (n, 3n) $^1$H & &
T4) T (p, D) D & &  
T6) T ($\alpha$, n) $^6$Li \\
 \\

h1) $^3$He (n, p) T & &
h3) $^3$He (n, $\gamma$) $^4$He & &
h5) $^3$He ($\alpha$, $\gamma$) $^7$Be \\
h2) $^3$He (n, D) D & &
h4) $^3$He (p, 2p) D & & \\
 \\

$\alpha$1) $^4$He (n, D) T & &
$\alpha$4) $^4$He  ($\alpha$, n)$^7$Be & &
$\alpha$5) $^4$He ($\alpha$, 2p) $^6$He \\
$\alpha$2) $^4$He (p, D) $^3$He & &
$\alpha$3) $^4$He ($\alpha$, p) $^7$Li & &
$\alpha$6) $^4$He ($\alpha$, np) $^6$Li \\ 
& &  & &$\alpha$7) $^4$He ($\alpha$, D) $^6$Li \\
 \\
\end{tabular} 
\vspace{0.3cm} 
 
\noindent The data on the corresponding cross sections  were taken from the following sources:

\begin{itemize}
\item[{\bf H1})] Reaction cross-section from ENDF/B-VI.8 evaluations \cite{ENDF} valid for a projectile energy up to 150 MeV.
% \item[ ]
\item[{\bf D1-D5})] Data from ENDF/B-VII.0\cite{ENDF} valid for projectile energy up to 150 MeV.
% \item[{\bf d6}] This channel is very weak but we took into account, anyway. For this reaction we have not found evaluations, therefore we have used the experimental data \cite{EXFOR}.
% \item[ ]
\item[{\bf T1,T3,T4})] Data from ENDF/B-VII.1 evaluations\cite{ENDF} valid for projectile energy up to 20 MeV.
\item[{\bf T2})] Data from JEFF-3.1 evaluations\cite{JEFF}, valid for projectile energy up to 20 MeV.
\item[{\bf T5})] For this reaction we have not found any theoretical evaluation, therefore we have used the experimental data.\cite{EXFOR} From 10-100 keV in CM-system, we have taken the data from Canon et. al. (2002).\cite{pT1} For 8.34 and 13.6 MeV we got it from Calarco et. al. (1983).\cite{pT2}
\item[{\bf T6})] Again no evaluations, we have extracted the data from R. Spiger and T. Tombrello (1967).\cite{pT4} These data are valid up to 18 MeV.
% \item[ ]
\item[{\bf h1-h4})] Evaluations from ENDF/B-VII.0, which are valid up to 20 MeV. 
\item[{\bf h5})] No evaluations. Experimental data in the interval 0.02-3.13 MeV collected by different authors (experiments from 2008-2009).
% \item[ ]
\item[{\bf $\alpha$1})]  Data from R. Shamu and J. Jenkin 1964 around 22MeV \cite{na1}, and one point at 50MeV differential cross sections from Sagle et. al. (1991).\cite{na2}
\item[{\bf $\alpha$2})] Experimental data for 31 MeV from Bunch et. al. 1964,\cite{pa1} for 32 MeV, 40 MeV, 50 MeV and 52.5 MeV from Sagle et. al. (1991),\cite{pa2} for 85 MeV from Votta et. al. (1974),\cite{pa3} and for 200 MeV from Alons et. al. (1986).\cite{pa4}
\item[{\bf $\alpha$3-$\alpha$7})] Data taken from King et. al. (1977) for energies between threshold and 50 MeV.\cite{aa1} For energies higher than 60 MeV, measurements from Glagola et. al. (1982)\cite{aa2} and fits from Mercer et. al. (2001) were used.\cite{aa3} In the work of King et. al. (1977), there are also data for the excited-state cross section of $^7$Li(0.478) production $\gamma$-ray line.
\end{itemize}

Finally, it is important to note that nuclear decay becomes a non-negligible effect over sufficiently long timescales. In particular, neutron decay plays a major role in our problem due to its high abundance and relatively short lifetime, $\tau_n =885.7$ s.\cite{PDG} We have taken into account neutron decay in our calculations.

\section{Network solutions}\label{subsec:1}

After calculating all the $<\sigma\,v>_{ij}^k$, we are ready to solve the system of differential equations Eq.(\ref{eq:abund}) for each species in the network. Our network contains a relatively large number of nuclear species (around 270). Therefore the system of differential equations is too complicated to be solved analytically, and must be solved numerically. We used the Matlab\textsuperscript{\textregistered}'s \emph{``ode15s''} solver, which is an implementation of the Klopfenstein-Shampine family of Numerical Differentiation Formulas (NDF) of orders 1 to 5.\cite{matlab}

The numerical solver does not guarantee a physically meaningful solution, we then included some constraints to be checked in every solution step. The constraints originate from conserved quantities such as the number of nucleons (or baryon number conservation), electric charge, etc. One of the most important constraints is baryon number conservation, which can be written as

\begin{equation}
\label{eq:barion_number}
\dot{\rho}= 0~~\Rightarrow~~\sum \dot{X}_i \equiv 0~~\Leftrightarrow~~\sum A_i\dot{n}_i\equiv 0,
\end{equation}

\noindent where $\rho\equiv\sum A_in_i$ is the number density of nucleons. This condition ensures that the number of nucleons remains constant even if, for instance, neutrons decay.

\subsection{Evolution of the plasma chemical abundance and the $\gamma$-ray emission}

We present the solution of the nuclear network for some simple cases. As a first example, we considered a unit volume of plasma at two fixed-temperatures, 1 MeV and 10 MeV. The nucleon number density for this plasma is $\rho=10^{18}~\rm{cm^{-3}}$. We studied three special cases, each of them differing only in the initial plasma composition. The initial mass abundance compositions for each case are:\\

\noindent 1) $X_p=0.7$ and $X_{^{12}\rm{C}}=0.3$\\
2) $X_p=0.7$ and $X_{^{28}\rm{Si}}=0.3$\\
3) $X_p=0.7$ and $X_{^{56}\rm{Fe}}=0.3$\\

\begin{figure}[!htp]
\begin{center}
 
\begin{tabular}{ll}
\includegraphics[scale=0.4]{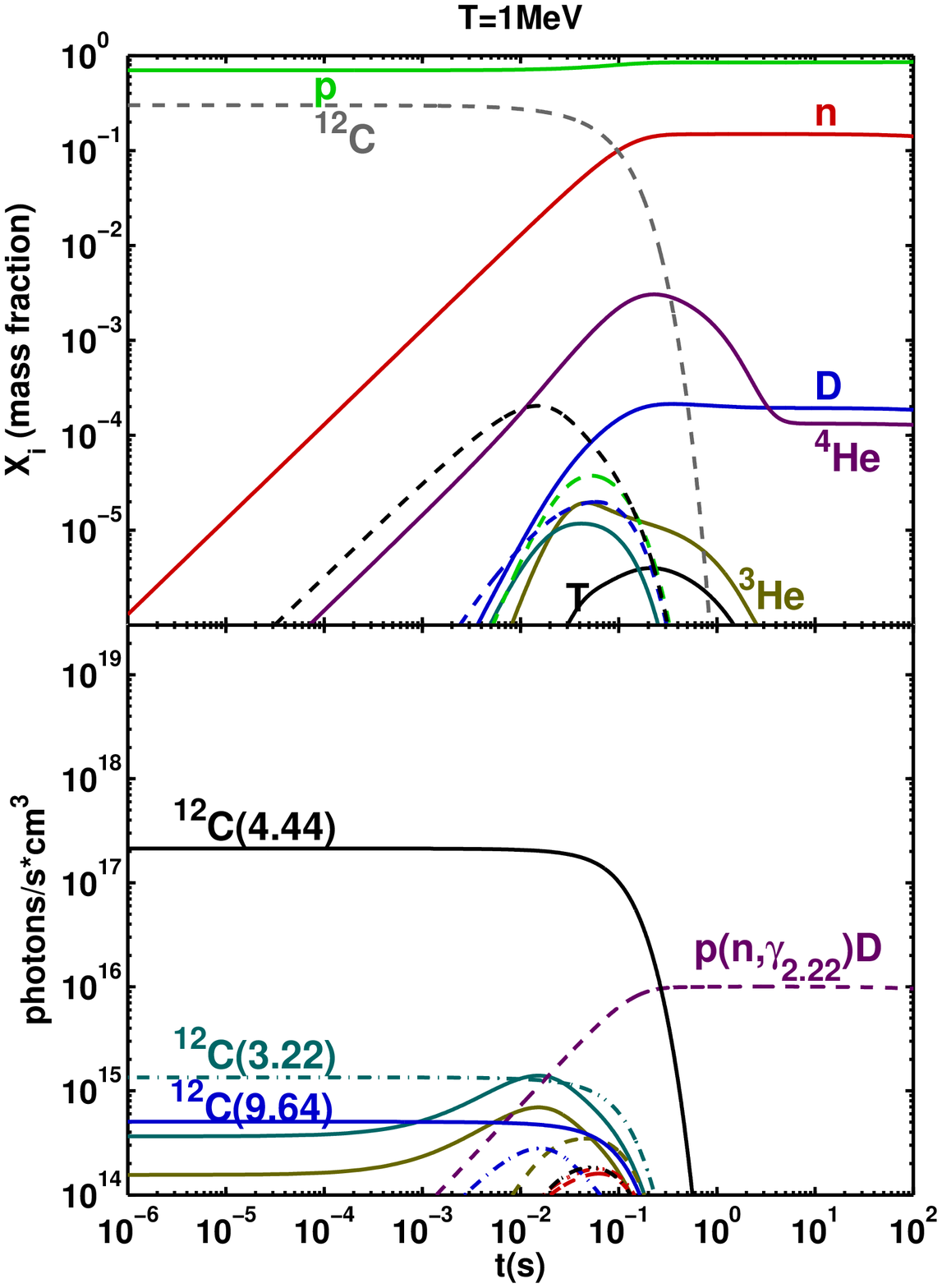} & 
\includegraphics[scale=0.4]{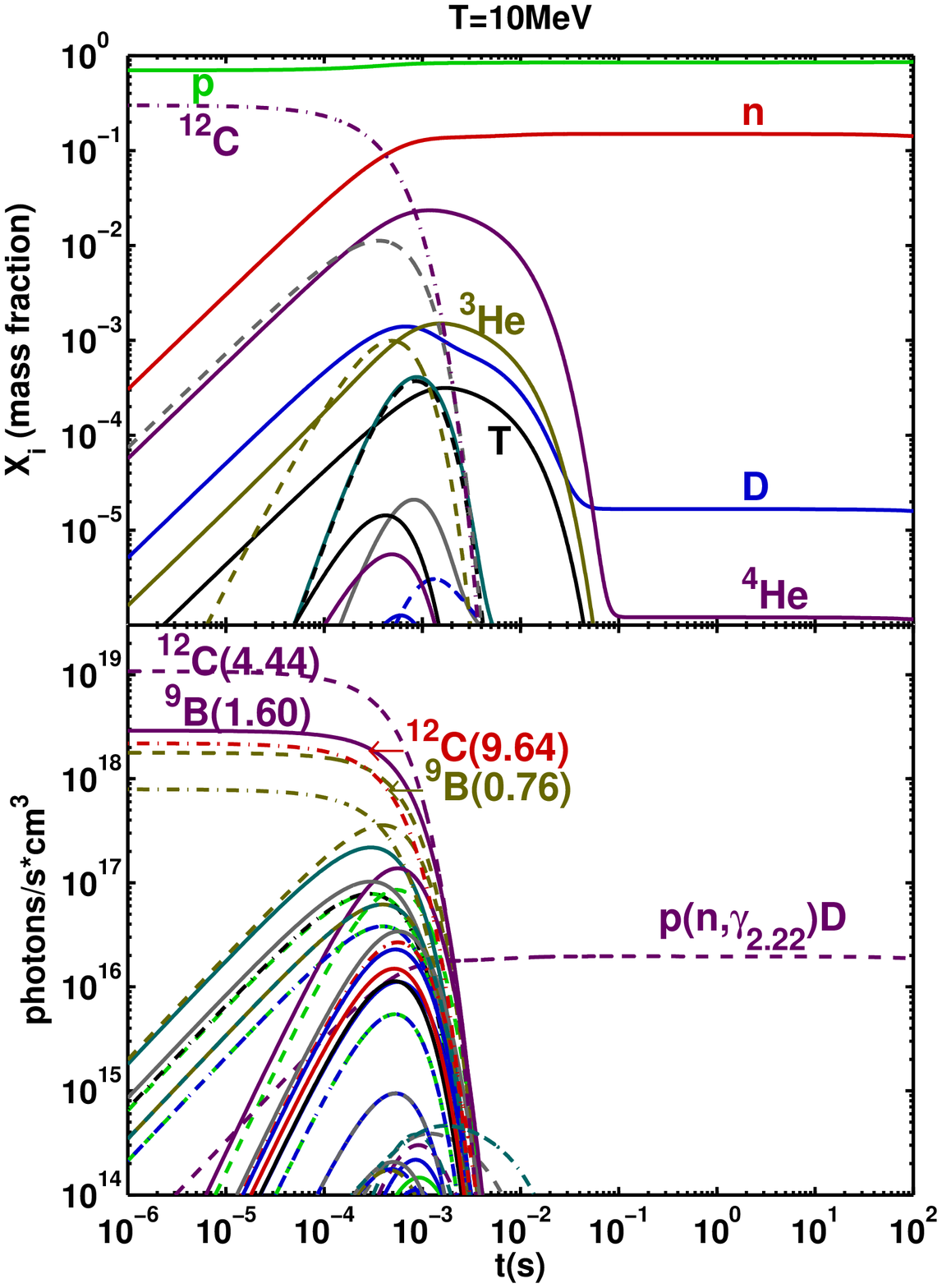}\\
\includegraphics[scale=0.4]{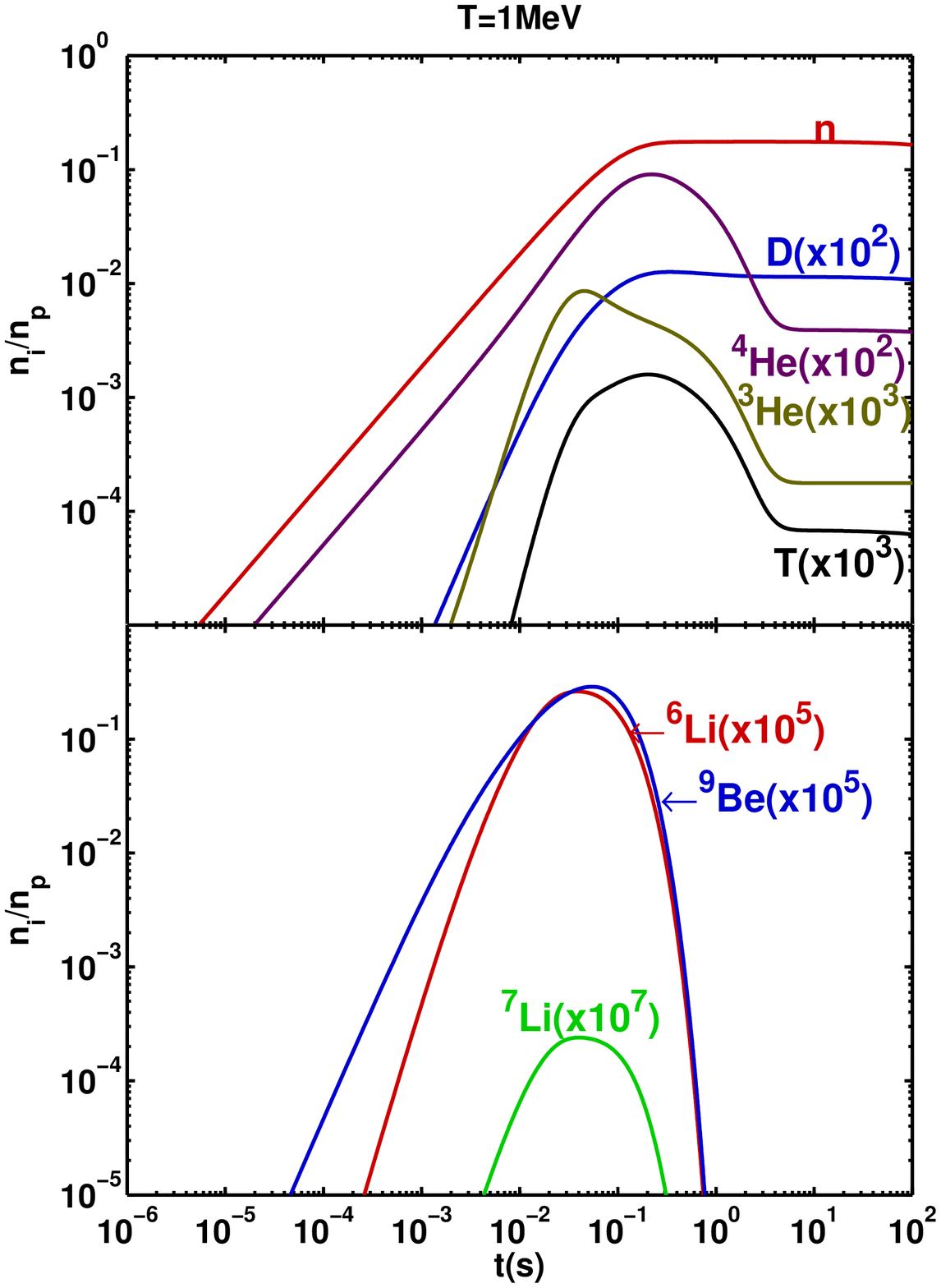} & 
\includegraphics[scale=0.4]{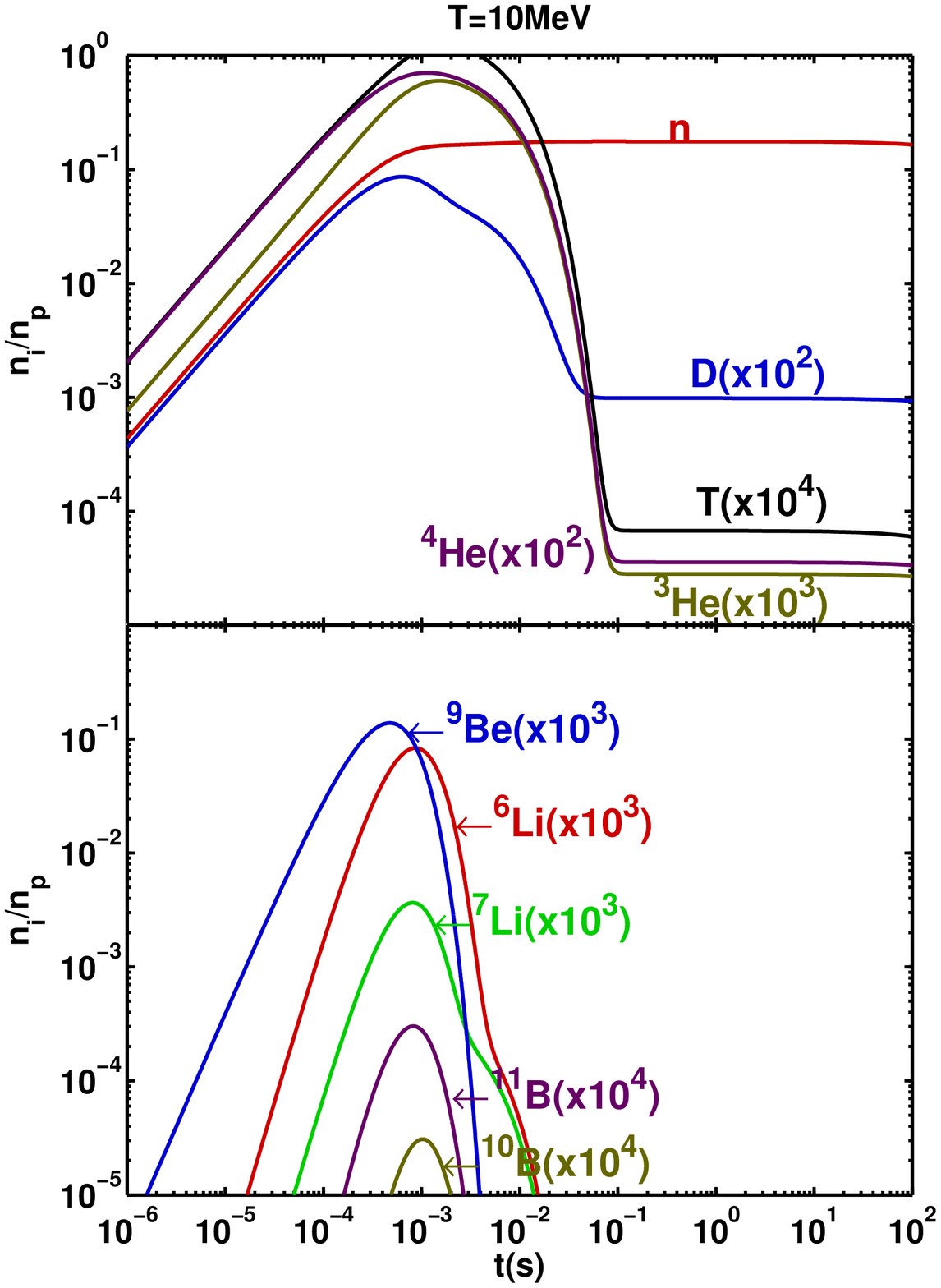}\\
\end{tabular}

\end{center}
\caption{Temporal evolution of mass-abundances and $\gamma$-ray line emission for a plasma with initial composition of $70\%$ protons and $30\% ~^{12}$C, for temperatures 1 MeV (left) and 10 MeV (right). In each case, the bottom panels show in particular the evolution of the abundance of light elements up to $^{11}$B, multiplied by the coefficient in brackets for comparison.}
\label{fig:c}
\end{figure}

\begin{figure}[!htp]
\begin{tabular}{ll}
\includegraphics[scale=0.4]{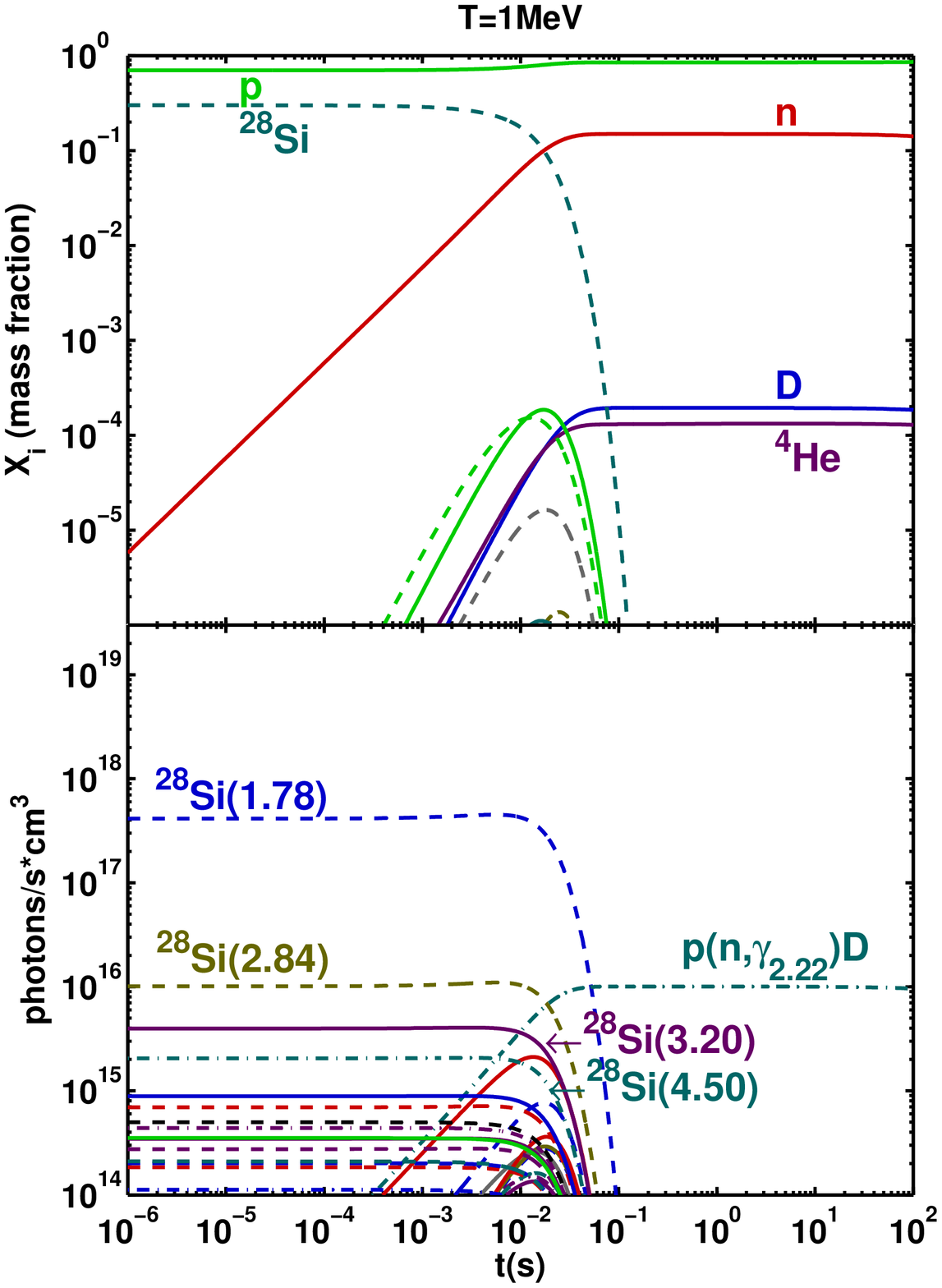} & 
\includegraphics[scale=0.4]{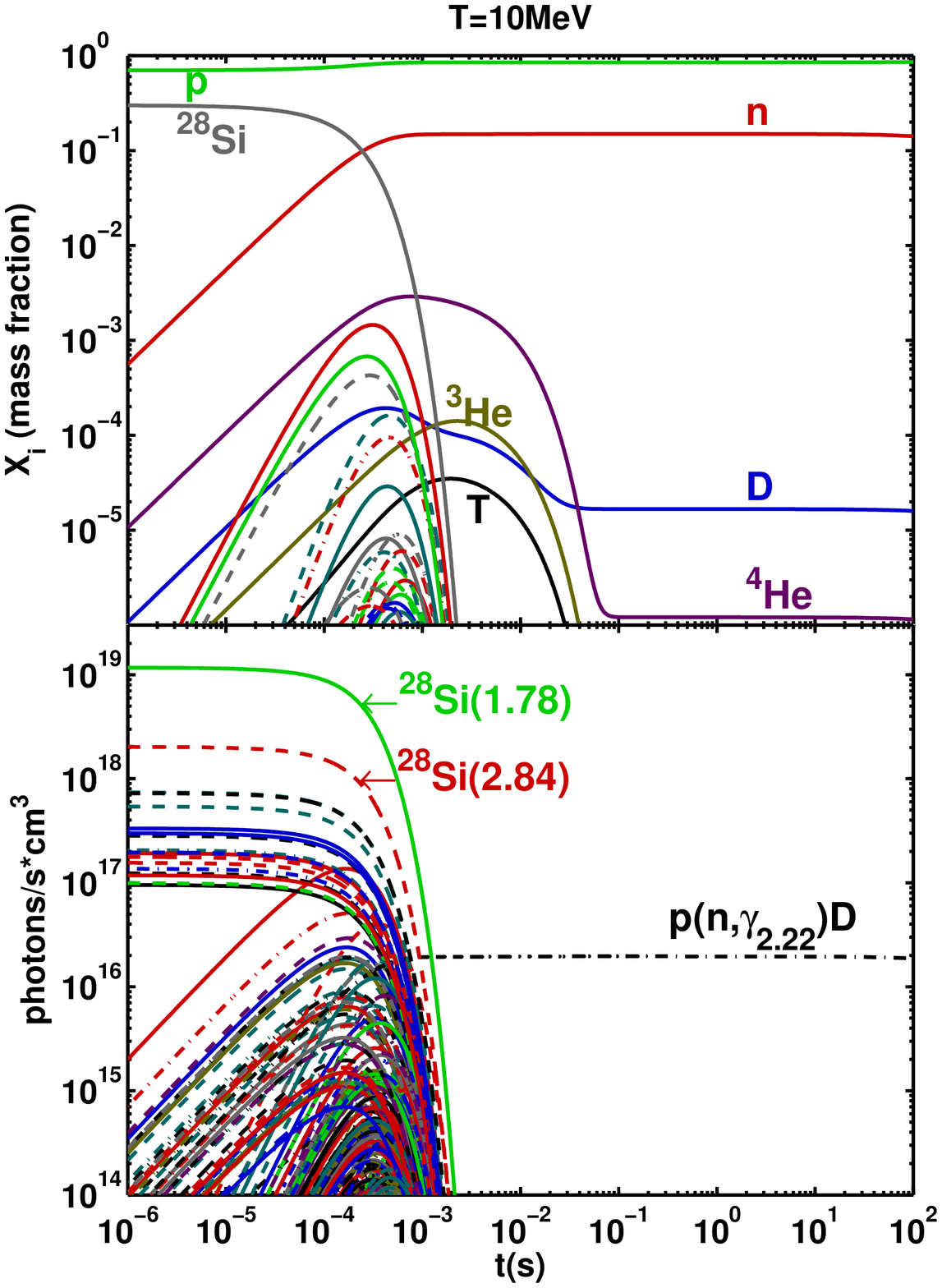}\\
\includegraphics[scale=0.4]{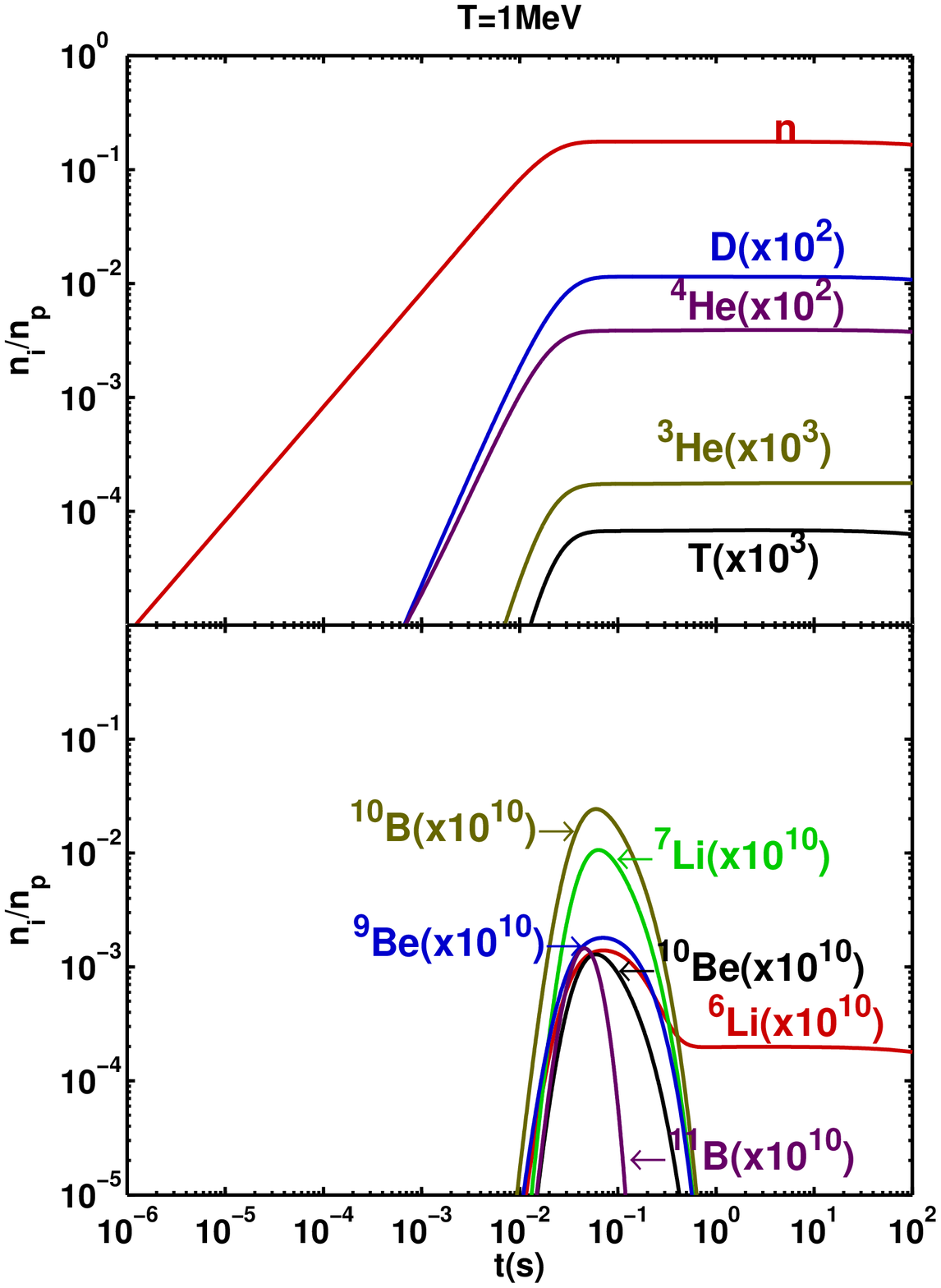} & 
\includegraphics[scale=0.4]{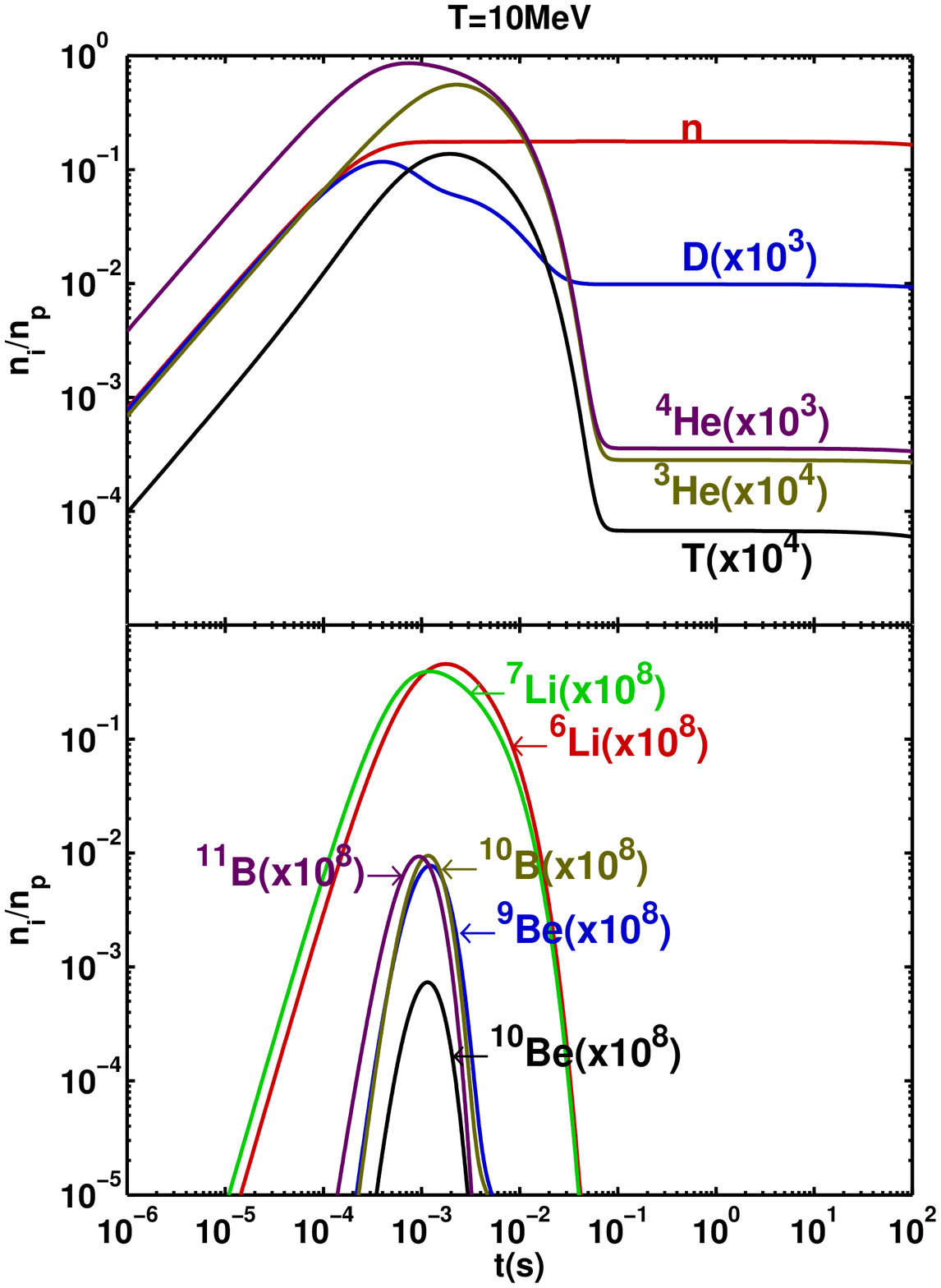}\\
\end{tabular}
\caption{Temporal evolution of mass-abundances and $\gamma$-ray line emission for a plasma with initial composition of $70\%$ protons and $30\% ~^{28}$Si, for temperatures 1 Mev (left) and 10 MeV (right). In each case, the bottom panels show in particular the evolution of the abundance of light elements up to $^{11}$B, multiplied by the coefficient in brackets for comparison.}
\label{fig:si}
\end{figure}

\begin{figure}[!htp]
\begin{tabular}{ll}
\includegraphics[scale=0.37]{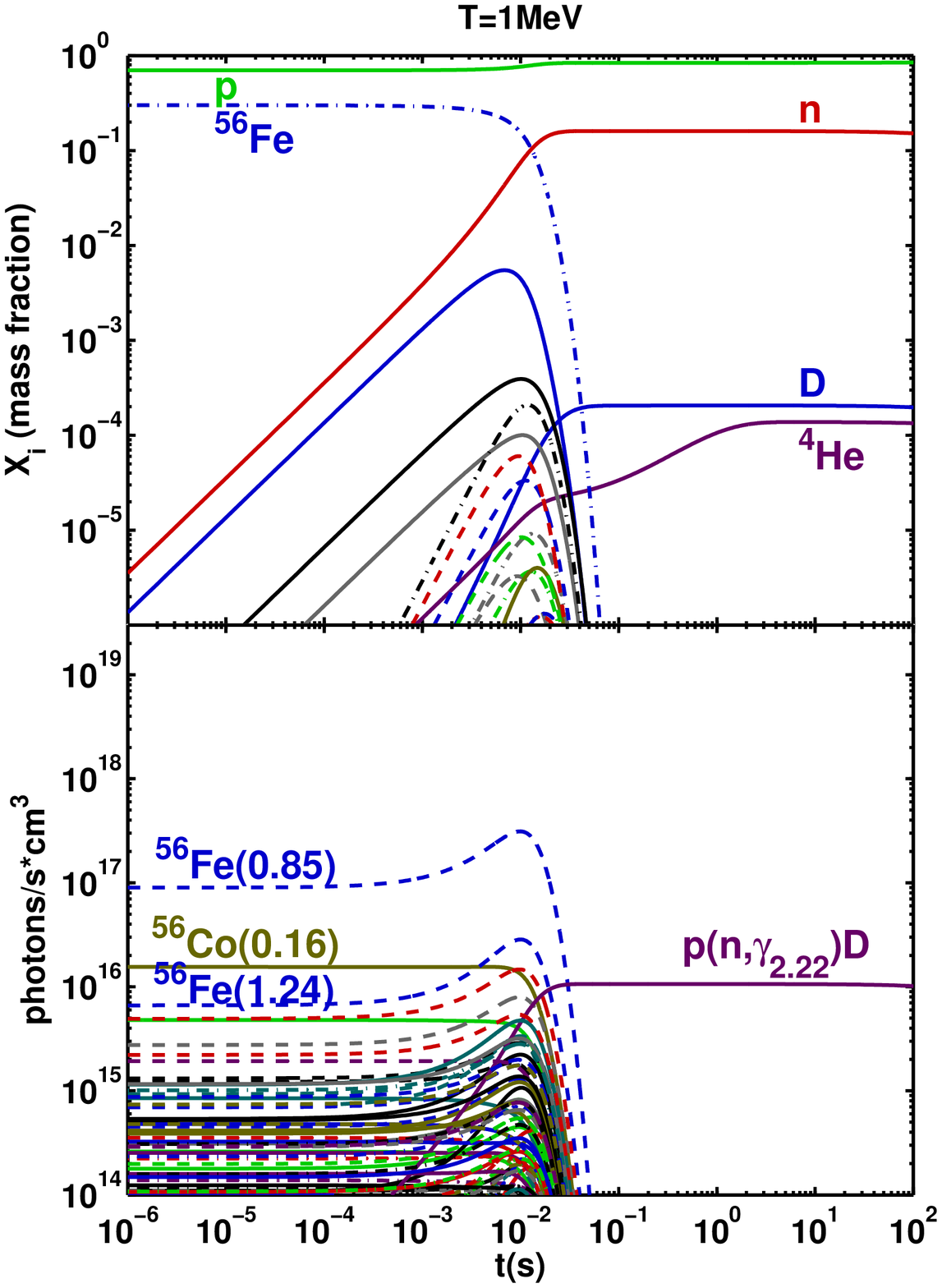} & 
\includegraphics[scale=0.37]{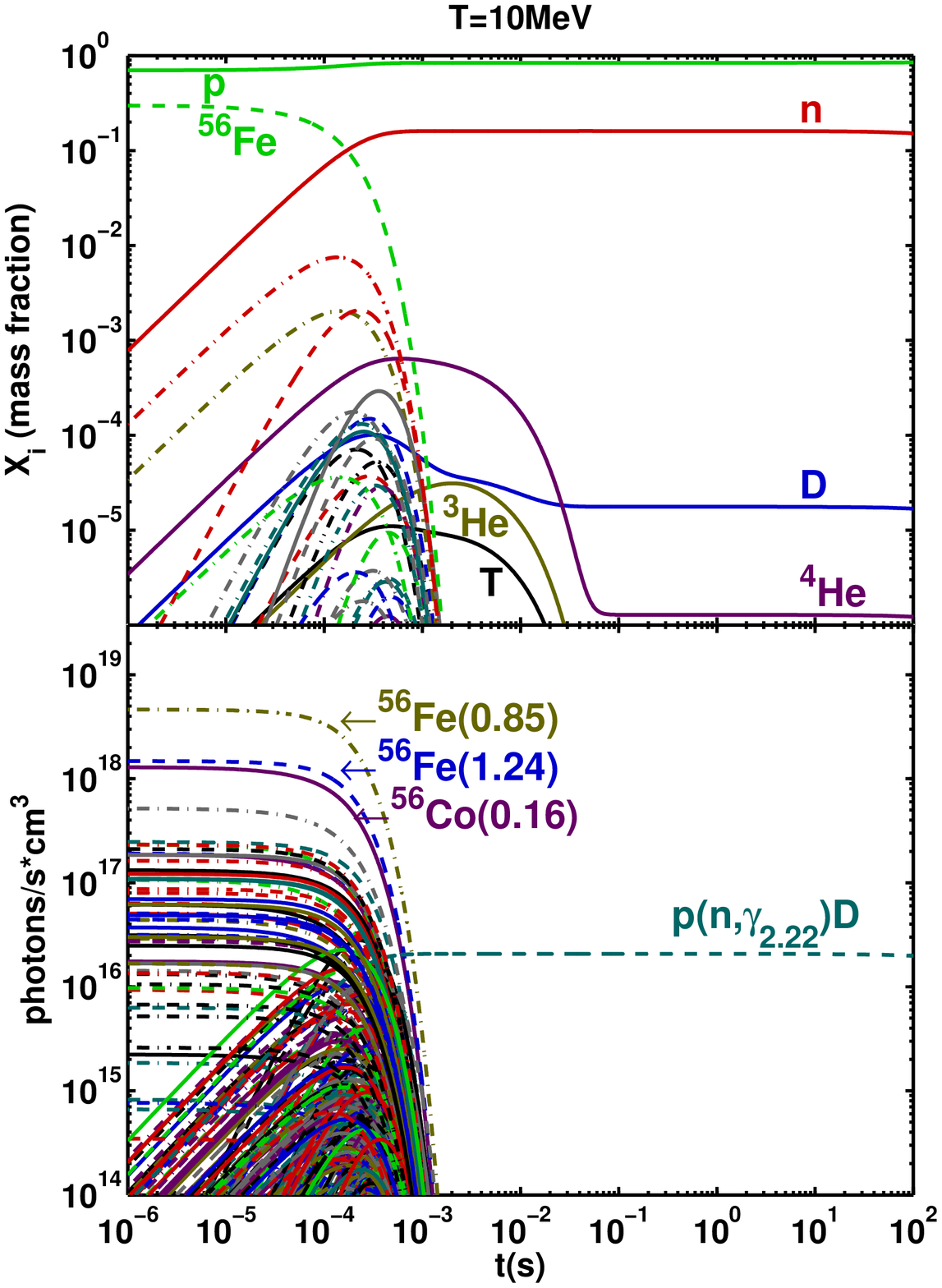}\\
\includegraphics[scale=0.37]{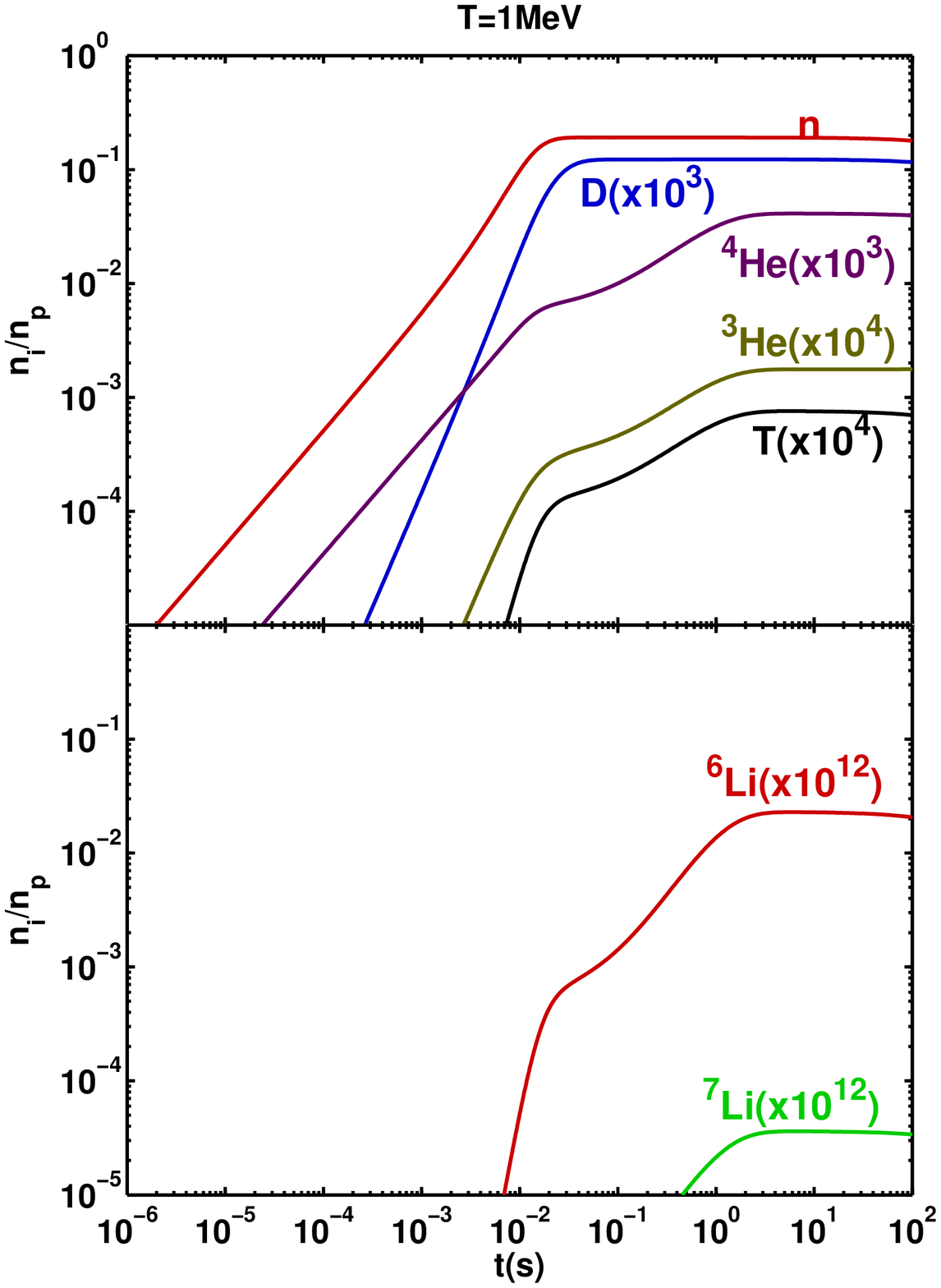} & 
\includegraphics[scale=0.37]{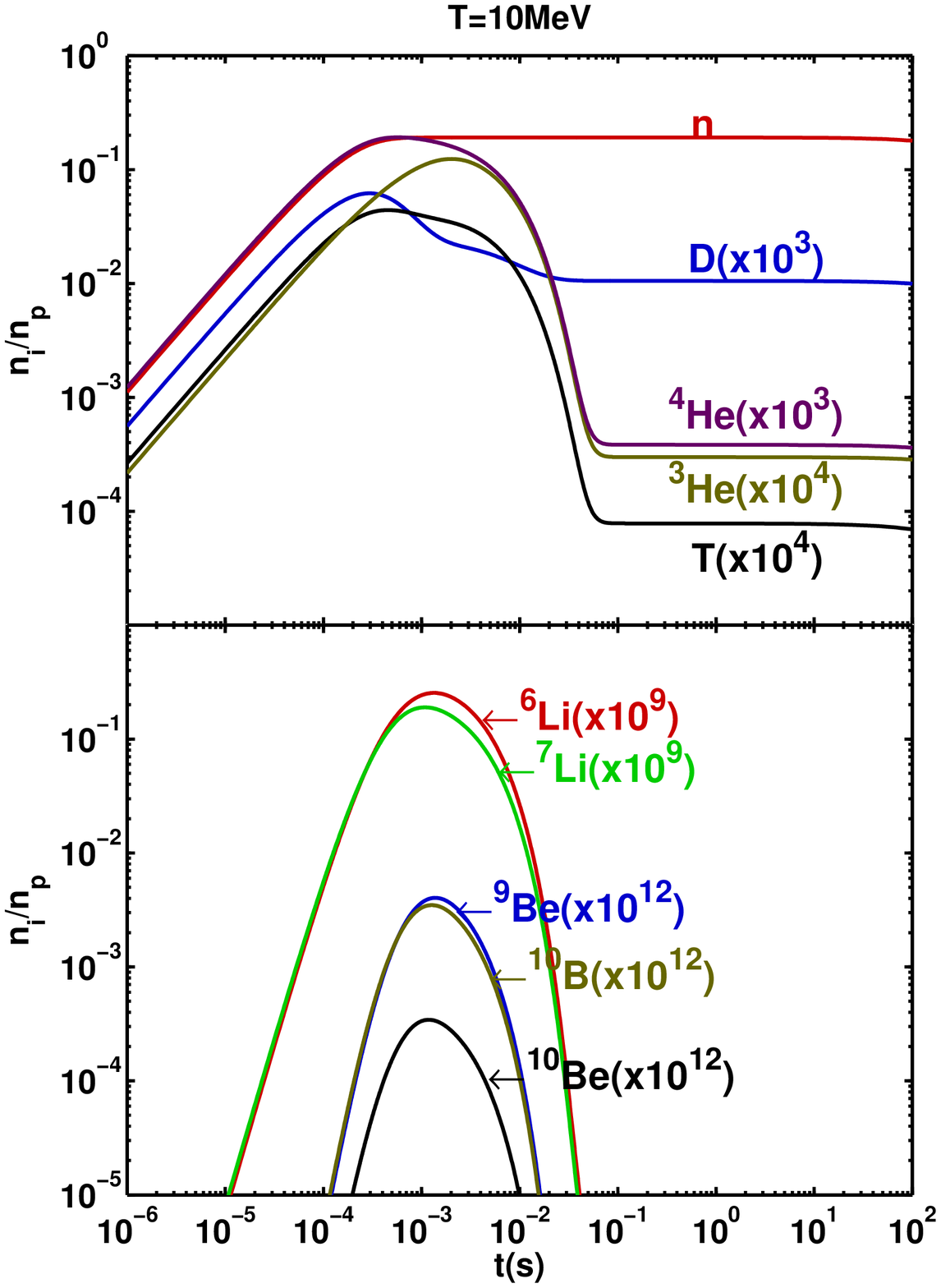}\\
\end{tabular}
\caption{Temporal evolution of the mass-abundances and $\gamma$-ray line emission for a plasma with initial composition $70\%$ protons and $30\% ~^{56}$Fe, for temperatures 1 Mev (left) and 10 MeV (right). In each case, the bottom panels show in particular the evolution of the abundance of light elements up to $^{11}$B, multiplied by the coefficient in brackets for comparison.}
\label{fig:fe}
\end{figure}

\noindent This is equivalent to having plasma with $n_p=7\times10^{17}\rm{cm^{-3}}$ and $n_{^{12}\rm{C}}/n_p=3.6~10^{-2}$, $n_{^{28}\rm{Si}}/n_p=1.53~10^{-2}$, and $n_{^{56}\rm{Fe}}/n_p=7.7~10^{-3}$, respectively. We then evolved the plasma for a period of 100 s. The temporal evolution of the plasma composition and the $\gamma$-ray line emission is shown in Figs. \ref{fig:c} to \ref{fig:fe}. Further discussion on these results is presented in Section \ref{discus}.
%
%\begin{figure}[!htp]
%\begin{tabular}{ll}
%\includegraphics[scale=0.5]{./plots/sol/T=1MeVHe4.eps} & 
%\includegraphics[scale=0.5]{./plots/sol/T=10MeVHe4.eps}\\
%\end{tabular}
%\caption{The time evolution of mass abundances and gamma ray lines emission for a plasma with initial composition of $70\%$ protons and $30\% ~^{4}$He, for temperatures 1 and 10 MeV left and right respectively.}
%\label{fig:he}
%\end{figure}
%
%\newpage
%

\subsection{The plasma's baryonic component $\gamma$-ray spectra}\label{subsec:2}

We calculated the $\gamma$-ray spectrum result only of baryonic interactions. For high temperature and low density plasmas, the dominant baryonic $\gamma$-ray emission processes are nuclear reactions and proton-neutron ($p$-$n$) Bremsstrahlung, see Ref.(\refcite{AS84}). Proton-neutron Bremsstrahlung and nuclear reactions such are the capture processes $^1$H$(n,\gamma)$D, D$(n,\gamma)$T, D$(p,\gamma)^3$He, and T$(p,\gamma)^4$He, give rise to the continuum component of the spectrum. The discrete component is purely due to the de-excitation of nuclei, that decay into different isomeric states leading to the emission of specific $\gamma$-ray lines.

We define $\Phi_\gamma$ as the gamma-ray emissivity per unit energy (or unit frequency). If $\dot{n}_\gamma$ is the number of $\gamma$-rays emitted per unit time and volume at energy $E_\gamma$, then the emissivity can be written as follows:

\begin{equation}
 \Phi_\gamma(E_\gamma) = E_\gamma~\frac{d\dot{n}_\gamma(E_\gamma)}{dE_\gamma}.
\end{equation}

Gamma-rays are directly created in capture reactions of the form $1+2\to3+\gamma$, such as $n+p\to \rm{D} +\gamma$. The emissivity for a capture reaction is given by

\begin{equation}
\label{eq:emissivity_capture}
\Phi_\gamma(E_\gamma) =n_1\,n_2\; c\;\frac{m_3}{m_1+m_2}\,\sqrt{\frac{8}{\pi\,\mu\,c^2\,(k_BT)^3}}\;E_\gamma\;I(E_\gamma),
\end{equation}

\noindent where 

$$
I(E_\gamma) =\int dE\,\frac{E\,(E+Q)\,\sigma(E)}{E_\gamma^2}\,\exp{\left[-\frac{m_3^2c^2}{2(m_1+m_2)k_BT}\left(1-\frac{E+Q}{E_\gamma}\right)^2\,-\,\frac{E}{k_BT}\right]}.
$$

\vspace{0.2cm}

\noindent Here $n_1$, $n_2$ are the number densities of species $1$ and $2$, $\mu$ is the reduced mass of the interacting system (nuclei 1 and 2), $E_\gamma$ is the $\gamma$-ray energy in the LAB-frame, $E$ is the collision energy in the center-of-mass frame, and $\sigma(E)$ and $Q$ are the cross section and the $Q$-value of the reaction, respectively.

The recoil of nuclei $3$ is neglected in Eq.(\ref{eq:emissivity_capture}), i.e. terms $\propto E_\gamma/2m_3$ are not included. The energy of the interacting particles is $\sim k_B\,T$, so most of the photons will be produced with an energy of the order of $E_\gamma\sim Q+k_B\,T$. Thus $(Q+k_B\,T)/m_3 < 10^{-2}$ and the approximation is justified.

The discrete gamma-ray spectrum is, as we have already mentioned, the result of the de-excitation of nuclei. The excited nuclei have a Maxwellian velocity distribution, therefore the emitted $\gamma$-ray lines will be Doppler-broadened. The Doppler broadening of a $\gamma$-ray line at energy $E_\gamma^0$ emitted by a nuclei of mass $m$ in a plasma at temperature $T$ is

\begin{equation}
\label{eq:doppler-broadening}
\Gamma_D=\sqrt{\frac{8\ln(2)\,k_BT}{mc^2}}\;E_\gamma^0.
\end{equation}

\vspace{0.2cm}

In a high-temperature plasma $\Gamma_D$ is much larger than the natural width $\Gamma_0$ for almost all the strongest lines. This simplifies the broadening to a simple Gaussian profile. If the energy of the photon emitted by the nucleus at rest is $E_\gamma^0$, and if $\dot{n}_\gamma$ is the number of photons with energy $E_\gamma^0$ that are emitted per unit time and volume, then the emissivity can be written as

\begin{equation}
\Phi_\gamma(E_\gamma) =\frac{E_\gamma^0\,\dot{n}_\gamma}{\sqrt{2\pi\sigma_G^2}}\;\exp{\left[-\frac{(E_\gamma-E_\gamma^0)^2}{2\sigma_G^2}\right]},
\end{equation}

\vspace{0.2cm}

\noindent where the Gaussian broadening $\sigma_G$ is

$$\sigma_G = \sqrt{\frac{k_BT}{mc^2}}\;E_\gamma^0.$$

\vspace{0.2cm}

Finally, for the calculation of the emissivity due to $n$-$p$ Bremsstrahlung we employed the cross section together with the calculations from Ref.(\refcite{AS84}). 

We now show the spectra on three specific moments during the plasma evolution, for the $^{12}$C, $^{28}$Si and $^{56}$Fe solutions that we have already presented in Section \ref{subsec:1}.

 \begin{figure}[!htp]
\begin{tabular}{ll}
\includegraphics[scale=0.37]{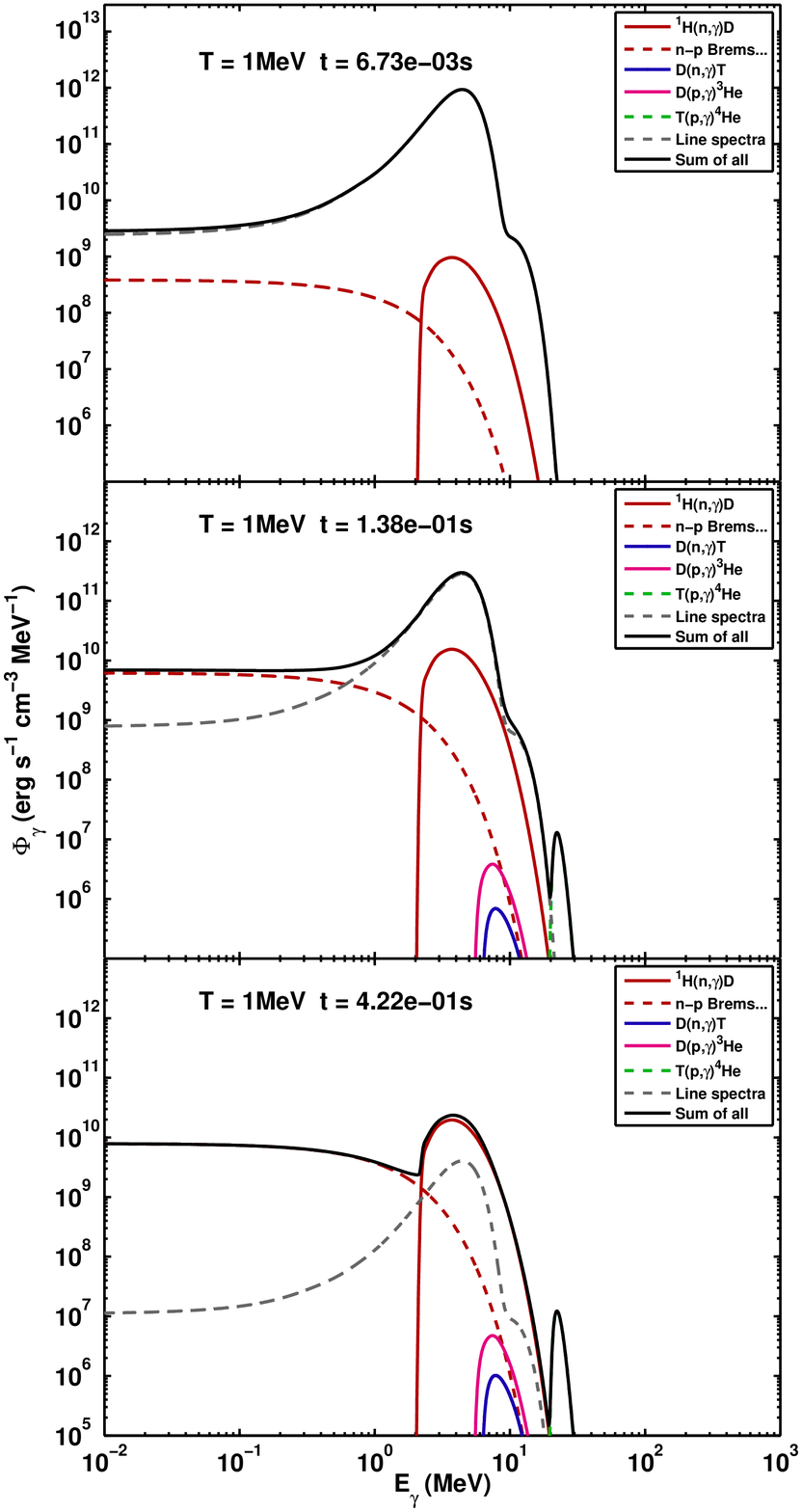} & 
\includegraphics[scale=0.37]{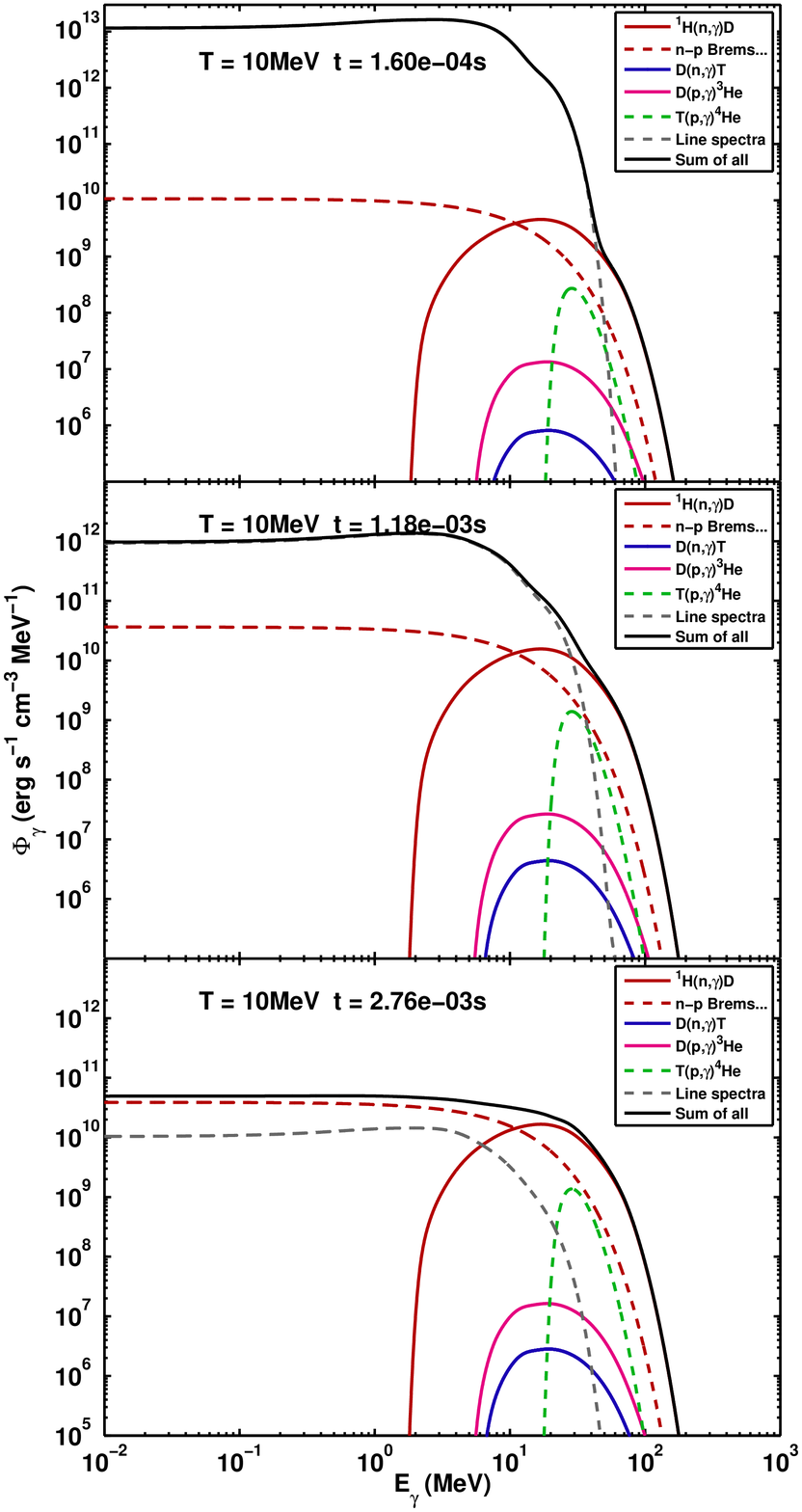}\\
\end{tabular}
\caption{Gamma-ray spectra of a plasma with initial composition of $70\%$ protons and $30\% ~^{12}$C, for temperatures 1 MeV (left) and 10 MeV (right).}
\label{fig:c-spec}
\end{figure}

\begin{figure}[!ht]
\begin{tabular}{ll}
\includegraphics[scale=0.37]{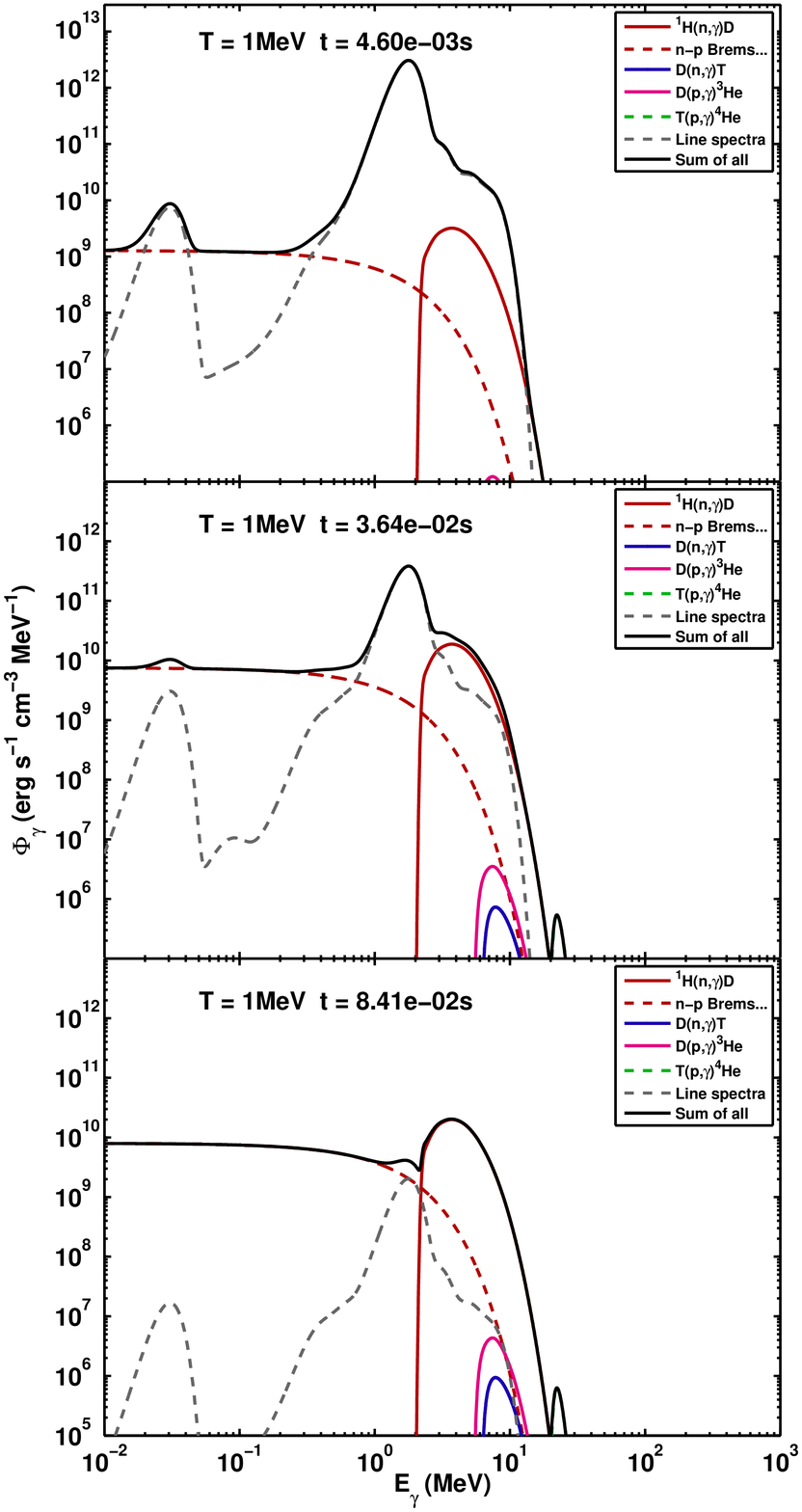} & 
\includegraphics[scale=0.37]{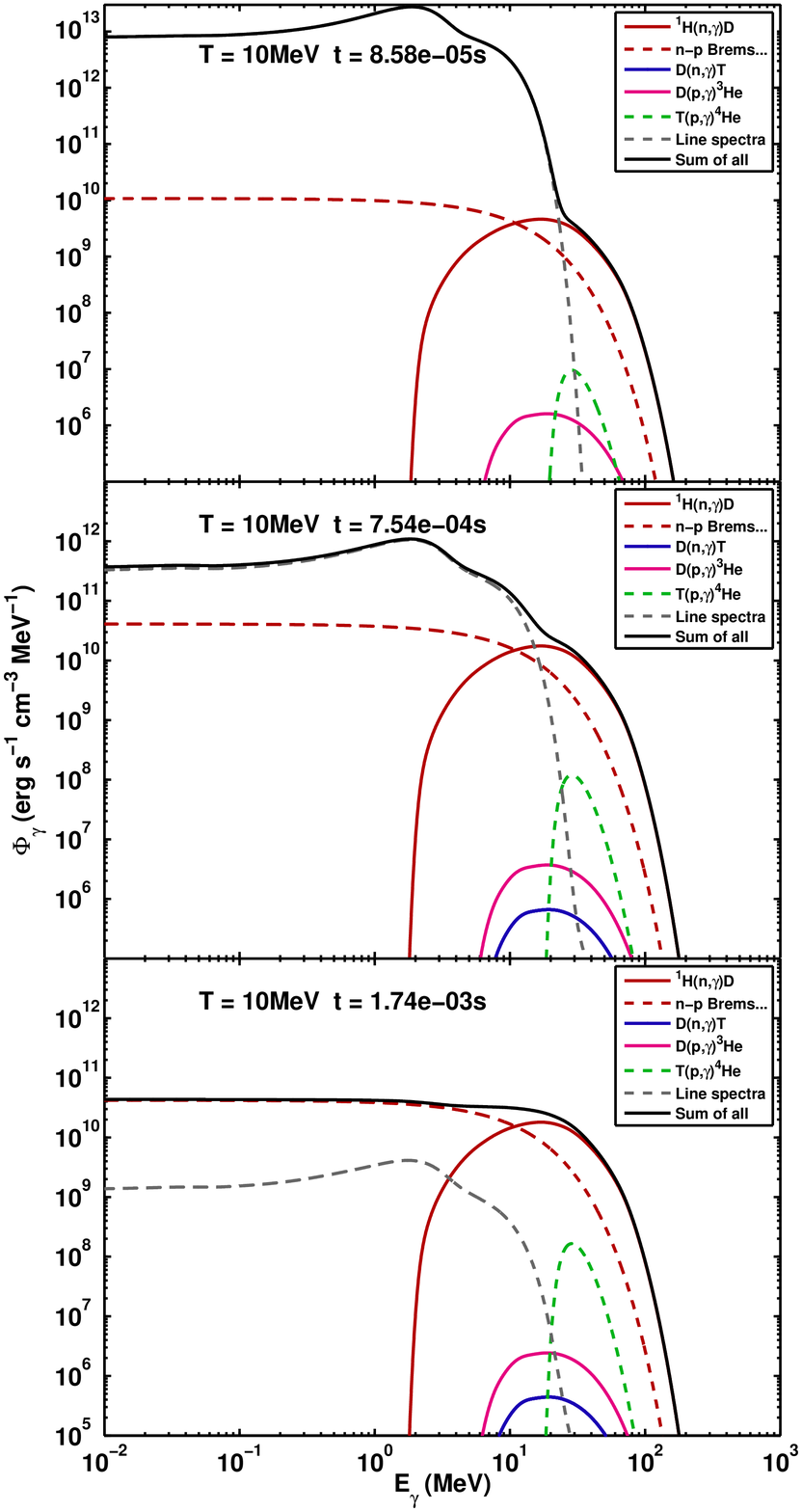}\\
\end{tabular}
\caption{Gamma-ray spectra for the plasma with initial composition of $70\%$ protons and $30\% ~^{28}$Si, for temperatures 1 MeV (left) and 10 MeV (right).}
\label{fig:fe-spec}
\end{figure}

\begin{figure}[!ht]
\begin{tabular}{ll}
\includegraphics[scale=0.37]{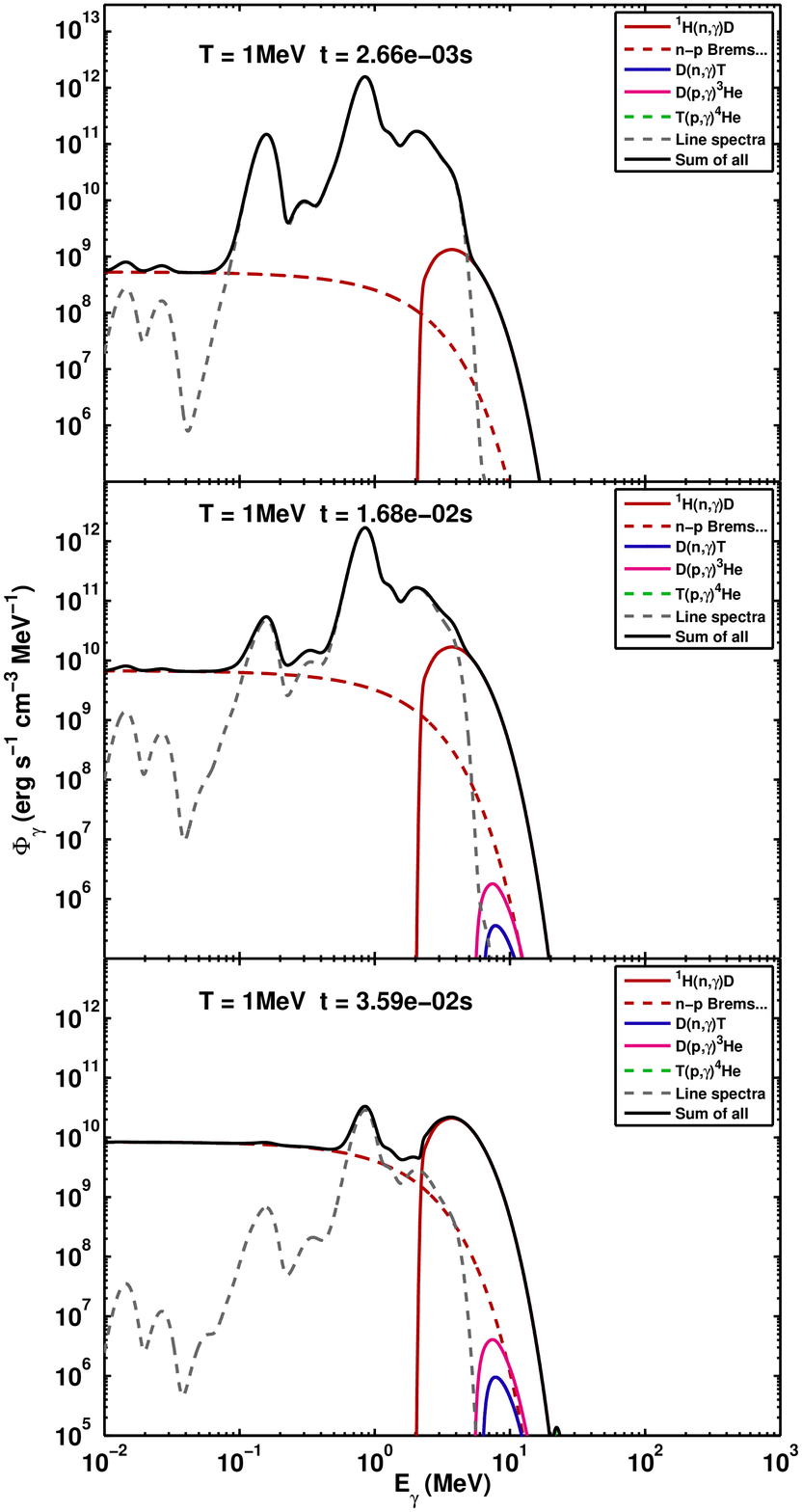} & 
\includegraphics[scale=0.37]{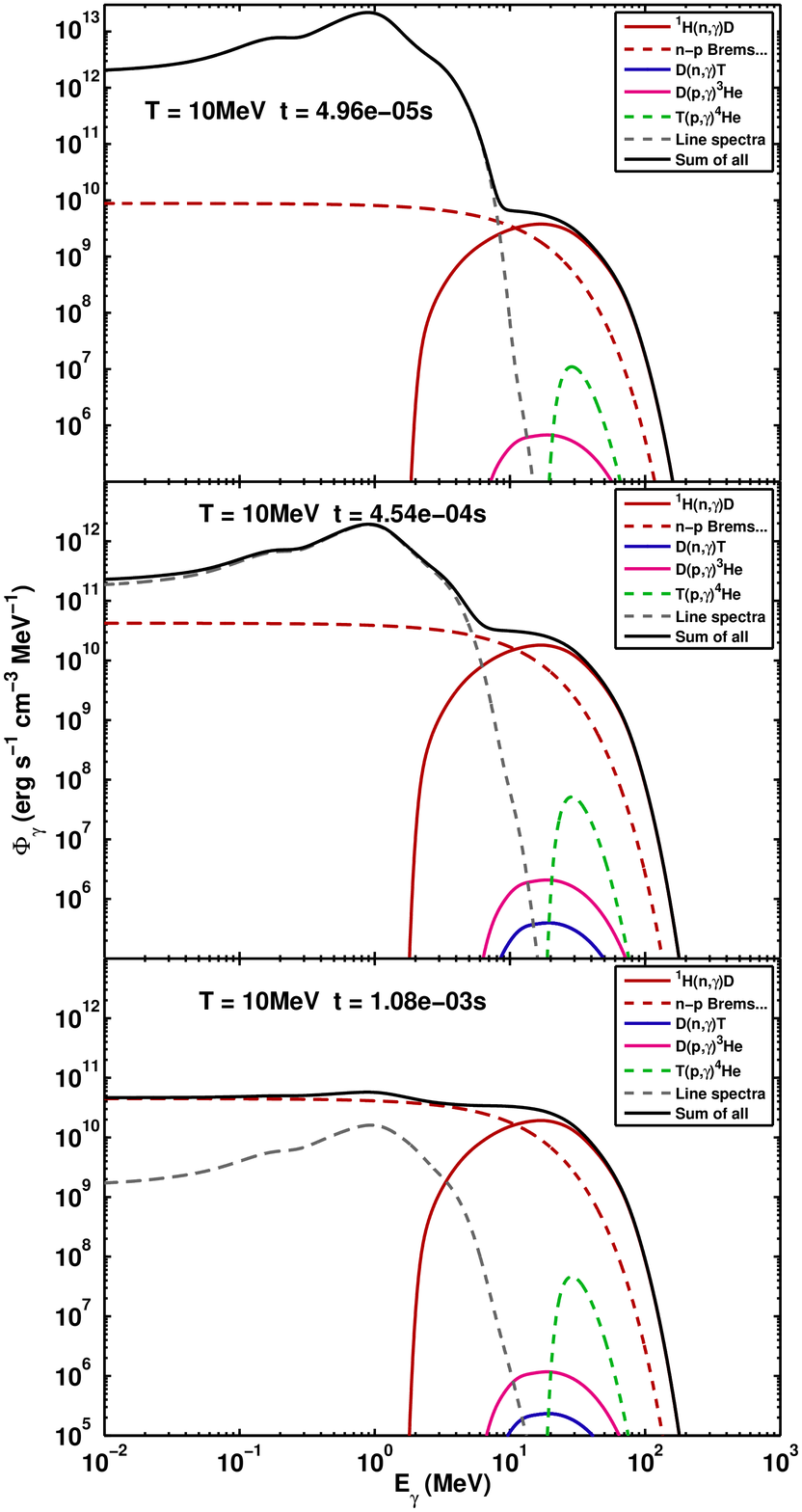}\\
\end{tabular}
\caption{Gamma-ray spectra for the plasma with initial composition of $70\%$ protons and $30\% ~^{56}$Fe, for temperatures 1 MeV (left) and 10 MeV (right).}
\label{fig:fe-spec}
\end{figure}

\newpage

\subsection{Other interesting examples} 

We now show, for comparison, two other interesting examples. The first corresponds to the case when the plasma temperature is assumed to be constant over the entire solution interval; in the second case the plasma cools on timescales comparable to nuclear destruction ones. 

To a first-order approximation, the result of the evolution of a low-density, high-temperature plasma over timescales larger than one second is a system formed by protons and neutrons. Since free neutrons decay, the plasma eventually consists purely of protons.

The other case considered here is that where the plasma cools ``instantly'': if $\tau_\Delta$ is the typical cooling time and $\tau_e$ the evolution timescale, then $\tau_\Delta\ll\tau_e$. Under this conditions many light and intermediate-mass elements can form. There is no physical impetus for this example; rather we performed this calculation to illustrate that the freezing out of the reactions with a change in temperature does indeed occur. However, a physically plausible scenario for such a change in temperature could be a plasma interacting with a fast shock. The same phenomenon could occur if the density and/or pressure are instantaneously lowered, such as in the case of free expansion.

We adopted a temporal dependence for temperature of the form $T=0.1+20\exp(-t/0.001)$~MeV to illustrate the freezing of reactions. Due to the added complexity and computational expenses, our network computes reaction rates down to $T=0.1$~MeV, and therefore we do not observe the total freezing.
\begin{figure}[!ht]
\begin{tabular}{ll}
\includegraphics[scale=0.5]{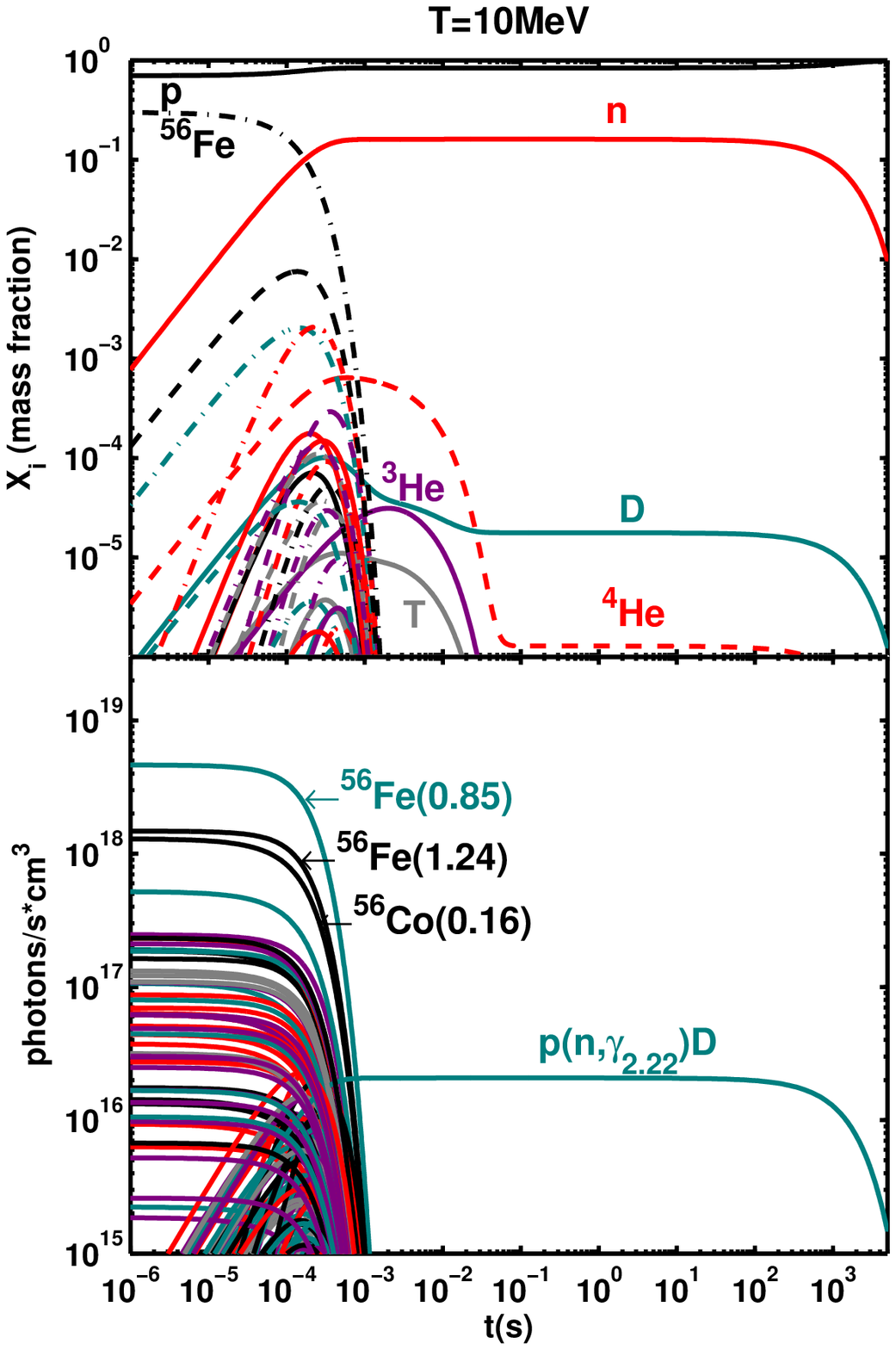}&
\includegraphics[scale=0.5]{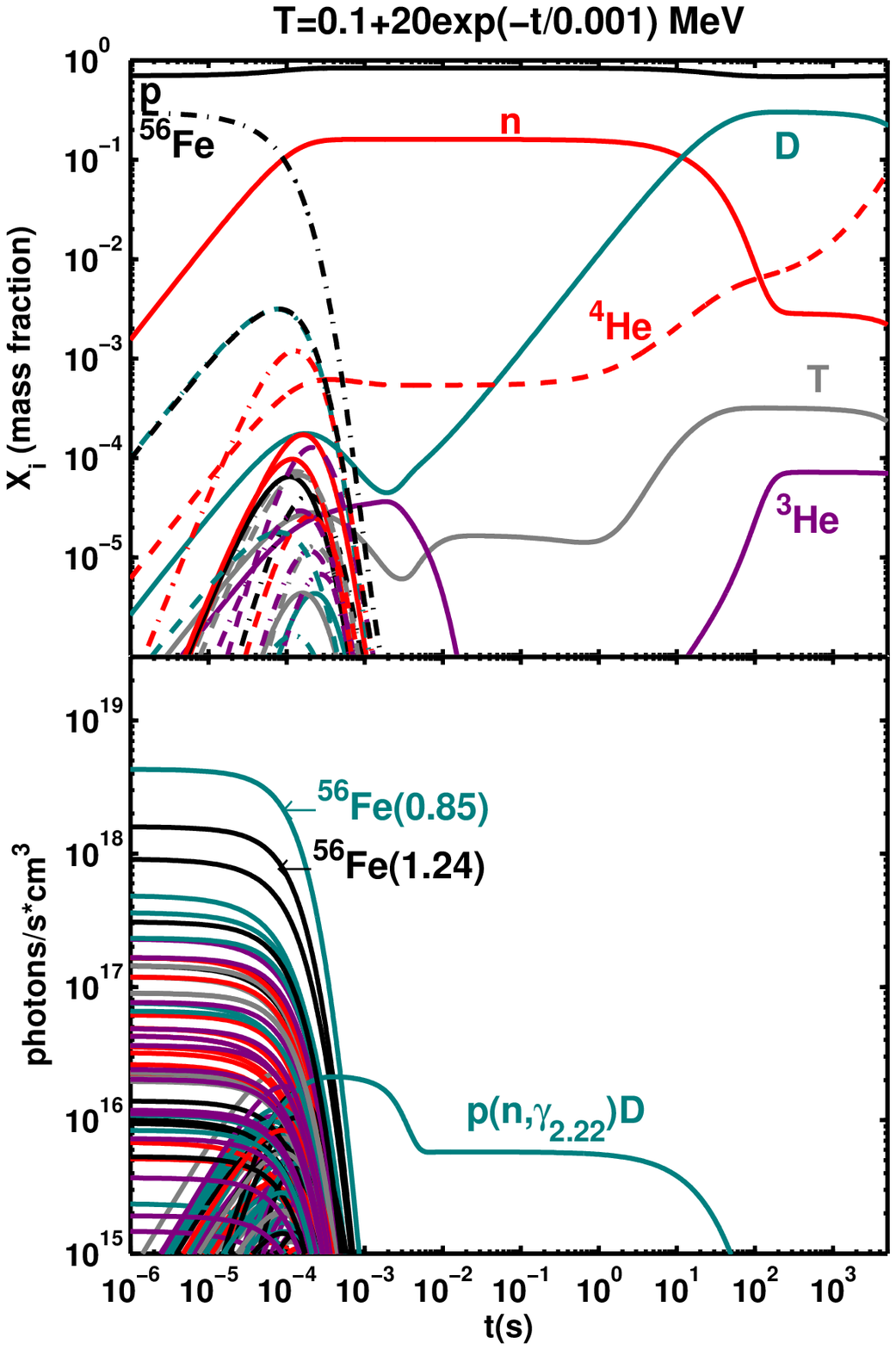}\\
\end{tabular}
\caption{The temporal evolution of the mass fraction abundances and the $\gamma$-ray line, emission for a plasma with initial composition of $70\%$ protons and $30\% ~^{56}$Fe, for constant temperature 10 MeV (left), and for time dependent temperature $T=0.1+20\exp(-t/0.001)$ MeV (right).}
\label{fig:time}
\end{figure}

\newpage

\section{Discussion}\label{discus}

As discussed above, we validated our network by calculating the mass abundance evolution of all the nuclei for some special cases. Figures \ref{fig:c} to \ref{fig:fe} show the temporal evolution of the mass fraction abundances and the number of photons of a given energy emitted per unit time and volume for the cases considered. Also shown is the number density ratio with respect to protons ($n_i/n_p$) for the light elements n, D, T, $^3$He, $^4$He and $^{6}$Li, $^{7}$Li, $^{9}$Be, $^{10}$Be, and $^{10}$B, $^{11}$B.

By comparing the plots on the left and on the right, we can draw some general inferences. It is clear that the value of the temperature has a strong impact on the characteristic timescales of the reactions. This is because the rates of almost all reactions increase with temperature, and correspondingly the nuclear abundances and the emission of $\gamma$-rays evolves much faster. Note from Figures \ref{fig:c} to \ref{fig:fe} that some intermediate elements (the result of break-up processes), become more abundant in certain periods of the evolution. At higher temperatures more interaction channels are opened, which leads to more $\gamma$-ray lines due to the de-excitation of nuclei. In the case when the plasma cooling timescale is such that all reactions freeze at a time when intermediate elements are relativity more abundant, this can result in higher-than-normal abundances of such elements at the end of the network's evolution, as shown in Fig. \ref{fig:time}.

The tendency of $T\geq1$~MeV plasmas is to destroy, very quickly, all nuclei heavier than $^4$He, and to a first approximation, we may conclude that on time scales $\tau\sim1$~s a proton-neutron plasma is formed. We also observe for times $t>1$~s that some small traces of deuterium, tritium, $^3$He and $^4$He are present. This is a result of the break-up processes and weak fusion reactions which take place on longer timescales. A result of the proton-neutron plasma formation is that all $\gamma$-ray line emission disappears. For timescales $\tau>1$~s, neutron capture by protons through the reaction $^1\rm{H}(n,\gamma)\rm{D}$, can produce deuterium and the emission of 2.22~MeV $\gamma$-ray photons. Due to the kinematics the 2.22~MeV emission line should be broadened and shifted -- by approximately $k_BT$ -- as is shown in Figs.~\ref{fig:c-spec} to Fig.~\ref{fig:fe-spec}.

Figures \ref{fig:c}-\ref{fig:fe} also show the temporal evolution of the number density ratio, $n_i/n_p$, of stable light elements with mass numbers up to $A=11$. The abundance of light elements such as Li, Be and B is practically negligible for a constant temperature and density plasma. Hydrogen and helium isotopes, however, are much more abundant due to proton and neutron equilibrium, provided that the system remains in equilibrium.

High temperature, low density plasmas have a characteristic $\gamma$-ray spectrum of baryonic origin. As we showed in Section \ref{subsec:2} the spectrum has two main components, a continuum and a discrete component. Because of the Doppler-broadening the lines overlap, the result being a ``continuum'' with some little bumps in the position of the strongest lines. The line spectrum dominates only on timescales shorter than nuclear destructions one; over longer timescales the dominant part of the spectrum comes from the proton-neutron plasma radiation, which is mainly $p$-$n$ Bremsstrahlung and $p+n\to D+\gamma$ capture.

The destiny of a long-lived (i.e., $t>\tau_n$), constant density and temperature plasma, is a purely protonic composition (cf. left panel of Fig. \ref{fig:time}). This is due to the $\beta$-decay of free neutrons into protons, and as a result no other element will remain in equilibrium. Ref. \refcite{AS84} suggested a ``neutron evaporation'' mechanism through which a fraction of the neutrons escape the plasma, so that high-temperature, low-density plasmas may be astrophysical neutron sources. The neutrons might be captured by surrounding (relatively) cold(er) material such as hydrogen and form deuterium, emitting 2.22~MeV $\gamma$-rays. The observation of this spectral line would serve to reveal the existence of the nearby neutron source.

Plasmas which for some reason are cooled to such an extent that reactions are frozen at the right moment in time, could produce a higher abundance of intermediate elements (and therefore light elements) as shown in the right panel of Fig. \ref{fig:time}. We conclude by suggesting that such a scenario could be physically plausible in situations where the plasma expands and cools, for example when being crossed by shock waves formed during a supernova explosion. Since supernova explosions are ubiquitous in the universe, this mechanism could, in principle, change cosmic abundances.

\newpage

\section*{Acknowledgments}
We thank Arjan Koning and Stephane Goriely for their helpful advice regarding TALYS. We also want to thank David Jones and Maxim Barkov for their very helpful suggestions.

\appendix
\section{Evaluations}\label{app:eval}

\subsection{Justification of the assumptions}\label{app:assum}

Here we intend to justify numerically the assumptions described in Section \ref{sec:cross}.

First of all, we have restricted ourself to a domain of low densities and high temperatures, i.e. $n\,\sim\,10^{18}\,\rm{cm^{-3}}$ and $T\ge1$ MeV. As we have already mentioned in Section \ref{sec:intro}, under such conditions break-up processes are dominant. Therefore, all heavy nuclei will be destroyed ``instantly'', resulting in a proton-neutron plasma with some traces of light nuclei such as D, T, $^3$He, $^4$He. 

Secondly, the astrophysical environments where these type of plasmas are supposed to exist are such that we are allowed to start our calculations with an initial plasma composition similar to solar or cosmic composition of elements.\cite{abund} 

The fact that all heavier nuclei are ``instantaneously'' destroyed, and that the initial composition is solar/cosmic (i.e. abundances of nuclei heavier than $^4$He are at least three orders of magnitude below), suggest that heavier nuclei than $^4$He will never become abundant. Below we attempt to quantify the importance of heavy nuclei interactions.

First of all, we want to illustrate the claim that break-up processes are the most important processes for $T>1$~MeV plasmas. In \ref{app:plot}, for some important nuclei, are compared total nonelastic with inelastic cross-sections, and their corresponding reaction-rates. In that work are also compared the total nonelastic and capture reaction cross sections together with the corresponding rates. The plots show that the rates of break-up processes (the difference between total nonelastic and inelastic) is bigger than excitation (inelastic) and capture reactions for $T>1$ MeV.

Secondly, by starting with a plasma composition similar to solar/cosmic chemical abundance, it is clear that the most abundant elements are hydrogen and helium, followed three to four orders of magnitude below by Oxygen, Neon, Nitrogen, Carbon, etc. Therefore, the major contribution will come from proton and alpha interactions due to their higher abundances.

To quantify it, let us look at the destruction timescales e.g. for the $^{16}$O nucleus, as the third most abundant element. Let us consider reactions of proton, alpha and $^{16}$O projectiles with $^{16}$O targets. The destruction reaction rates will be given by:

\begin{eqnarray}
\Gamma_{(p+^{16}O)}     &=& <\sigma_{(p+^{16}O)}\,v>\,n_p\\
\Gamma_{(\alpha+^{16}O)}&=& <\sigma_{(\alpha+^{16}O)}\,v>\,n_\alpha\\
\Gamma_{(^{16}O+^{16}O)}   &=& <\sigma_{(^{16}O+^{16}O)}\,v>\,n_{^{16}O}.
\label{eq:}
\end{eqnarray}

\noindent From Ref.~\refcite{abund} we get the ratios: $n_\alpha/n_p\approx 8\times10^{-2}$,  $n_{^{16}O}/n_p\approx 5.4\times10^{-4}$. 

To estimate the order of magnitude value of the reaction rate for the above reactions we simply considered constant cross sections with value equal to the peak value. From measurements in Ref. \refcite{p016} - \refcite{016}, we get these peak values: $ \sigma_{(p+^{16}O)}\sim 0.5$ b (total nonelastic), $\sigma_{(\alpha+^{16}O)}\sim 0.9$ b (total nonelastic), and  $\sigma_{(^{16}O+^{16}O)}\sim 1$ b (it is of the order of the sum of all reaction cross sections). 

The reaction rate for constant cross section is $<\sigma\,v>\,=\,\sigma\,<v>$ where $<v>$ is the mean relative velocity, $<v>\sim \mu^{-1/2}$,  where $\mu$ is the reduced mass. Hence, the destruction reaction rates ratios are:

$$\frac{\Gamma_{(\alpha+^{16}O)}}{\Gamma_{(p+^{16}O)}}~~\approx ~~~\frac{\sigma_{(\alpha+^{16}O)}}{\sigma_{(p+^{16}O)}}\,~\sqrt{\frac{\mu_{(p,^{16}O)}}{\mu_{(\alpha,^{16}O)}}}\, ~~~\left(\frac{n_\alpha}{n_p}\right)~\sim~ \,10^{-1}$$
$$\frac{\Gamma_{(^{16}O+^{16}O)}}{\Gamma_{(p+^{16}O)}}\approx \frac{\sigma_{(^{16}O+^{16}O)}}{\sigma_{(p+^{16}O)}}\,\sqrt{\frac{\mu_{(p,^{16}O)}}{\mu_{(
^{16}O,^{16}O)}}}\,\left(\frac{n_{^{16}O}}{n_p}\right)\sim \,10^{-4}$$

The last estimate shows that neglecting interactions with projectiles different from protons and alphas will cause an error $$\frac{\Gamma_{(^{16}O+^{16}O)}}{\Gamma_{(p+^{16}O)}+\Gamma_{(\alpha+^{16}O)}+\Gamma_{(^{16}O+^{16}O)}}< 0.1\%$$

Then, with an accuracy better than $0.1\%$ we may neglect all nuclear reactions except those that involve protons and/or alphas. One must note here that due to the break-up processes, the number of neutrons increases very fast and they become very abundant. Therefore neutron interactions become important soon. After taking all these into account, we conclude that the most important nuclear interactions in high-temperature, low-density plasmas, are those reactions which involve protons, neutrons and alphas as projectiles. The interactions of heavy nuclei will never become important.

\subsection{Photodisintegration}\label{app:gamma}

As discussed in \ref{app:assum}, the most important projectiles are protons, neutrons and alpha particles. Plasmas produce photons through ion-nuclear reactions and electron Bremsstrahlung (see Section.~\ref{sec:cross}). The photon number density depends on the optical depth, i.e. electron density and the plasma's size. The photon number density per $E_\gamma$ energy interval ($dn_\gamma/dE_\gamma$), is given by:
\begin{equation}
\frac{dn_\gamma}{dE_\gamma} \approx \frac{d\dot{n}_\gamma}{dE_\gamma}\times \frac{R}{c}\times \max(1,\tau),
\end{equation}

\noindent where  $d\dot{n}_\gamma/dE_\gamma$ is the photon production rate per $E_\gamma$ energy interval, $R$ is the size of the plasma, $\tau$ is the optical depth and $c$ is the speed of light. The term $\max(1,\tau)$ takes into account the photon's time-escape delay due to propagation through an optically thick plasma ($\tau>1$).

The optical depth for a homogeneous plasma with number density $n$ and size $R$ is given simply by $\tau(E_\gamma)\,=\,n~R~\sigma_c(E_\gamma)$, where $\sigma_c(E_\gamma)$ is the Compton scattering cross-section (Klein-Nishina).

The photodisintegration reaction-rate will be given by:

\begin{equation}
\Gamma_{(\gamma,x)}\,=\,\int dE_\gamma\frac{dn_\gamma}{dE_\gamma}\sigma_{(\gamma,x)}(E_\gamma)\,c
\end{equation}
where $\sigma_{(\gamma,x)}(E_\gamma)$ is the photodisintegration cross section for a given nuclei. The cross sections for the important nuclei such as D, $^{12}$C, $^{16}$O, $^{56}$Fe, etc, generally show a threshold $E_\gamma >3$~MeV, and have a maximum around $7-20$~MeV.

By including all of the above, one obtains:

\begin{equation}\label{app:photo}
\Gamma_{(\gamma,x)}\,=\,R\,\int dE_\gamma\left(\frac{d\dot{n}_\gamma(E_\gamma)}{dE_\gamma}\right)\times\sigma_{(\gamma,x)}(E_\gamma)\times\max\left(1,\tau(E_\gamma)\right)
\end{equation}

Note that the photodisintegration reaction rate in formula \ref{app:photo}, includes two cases, one without optical depth and one with optical depth. The case without optical depth scales as $n_e^2\,R$ ($n_e$ is the electron number density), whilst the case with optical depth scales as $n_e^3\,R^2$. On the other hand ion-nuclear reaction rates scale as $n_i^2$ ($n_i$ is the ion number density). Thus, for an optically thin plasma, photodisintegration effect is mostly influenced by the size $R$ because $n_e\sim n_i$.
% 
% If one may wish to show the photodisintegration reaction-rate dependence from plasma's size and electron number density (if $n_e=n_p$):
% $$
% \Gamma_{(\gamma,x)}\,=\,(R\,n_e^2)\,\int dE_\gamma \left({\frac{d\dot{n}_\gamma(E_\gamma)}{dE_\gamma}}\right)_0\,\sigma_{(\gamma,x)}(E_\gamma)+$$
% $$+\,(R^2\,n_e^3)\,\int dE_\gamma \left({\frac{d\dot{n}_\gamma(E_\gamma)}{dE_\gamma}}\right)_0\,\sigma_c(E_\gamma)\,\sigma_{(\gamma,x)}(E_\gamma)  
% $$

In high temperature ($T_i>1$~MeV), low density, optically thin ($\tau < 1$) plasma, the electrons are heated due to Coulomb exchange with ions and are cooled due to Bremsstrahlung, which has a higher rate than heating. In the extreme case of electrons getting heated to high temperatures ($T_e>1$~MeV), pair-production processes become non-negligible, and thus, the plasma will cool down (see Ref.~\refcite{Bisnovatyi71} and Ref.~\refcite{Svensson84}).

The electron thermal relativistic Bremsstrahlung ({\it e-e} and {\it e-ion}) emissivity is given in Ref.~\refcite{Gould80}. The {\it p-n} capture and Bremsstrahlung thermal emissivities are calculated as described in Section.~\ref{subsec:2}.

We consider here an uniform, electrically neutral, spherically symmetric, optically thin plasma, with a size $R\sim10^6~\rm{cm}$ and number density $n\sim10^{18}~\rm{cm^{-3}}$.

Two important nuclei are considered, the $^{16}$O as the third most abundant cosmic/solar element after Hydrogen and Helium, and Deuterium  as the most abundant light element obtained after heavier nuclei are destroyed (see figures of Section.~\ref{subsec:1}). By using the formula \ref{app:photo}, we calculate the reaction rates for $^{16}$O$(\gamma, X)$ and D$(\gamma, n)p$ photodisintegration reactions. For the sake of comparison, we calculated separately the photodisintegration reaction rates of photons emitted by electrons and ions, and we include the $^{16}$O$(p, X)$ proton break-up. The results are shown in Fig.~8. We notice that reaction rates coming from electron Bremsstrahlung photons are a function of electron temperature, whereas the reaction rates coming from nuclei are a function of ion temperature.

\begin{figure}[!h]
\label{app:plot}
 \includegraphics[scale=0.5]{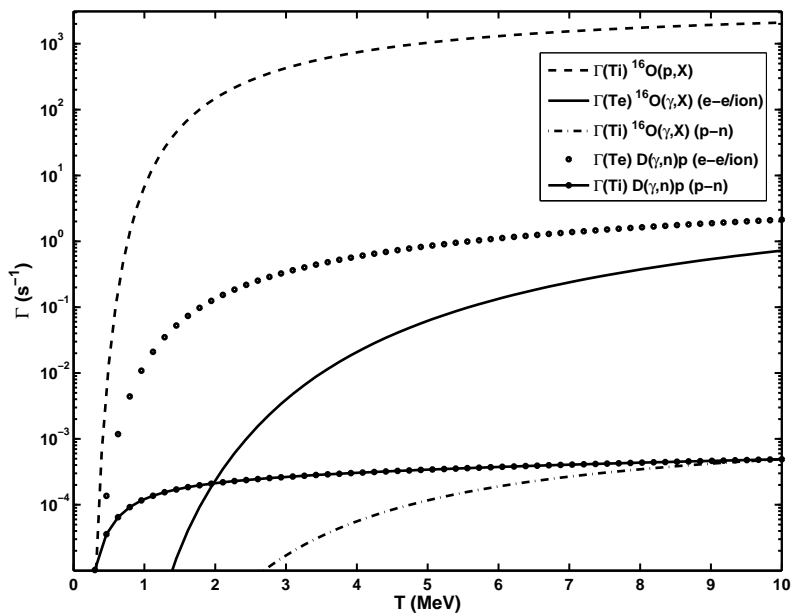}
\caption{The photodisintegration reaction rate for $^{16}$O and D nuclei as a function of temperature, for a plasma with size $R=10^6~\rm{cm}$, and $n=10^{18}~\rm{cm^{-3}}$. The full line ($^{16}$O) and the dotted curve (D) are the photodisintegration reaction rates of photons comming from electron Bremsstrahlung as a function of electron temperature ($T_e$). The dash-dotted ($^{16}$O) and the full-dotted (D) lines are the photodisintegration reaction rates of photons coming from nuclei as a function of ion temperature ($T_i$). For comparison the dashed curve is the proton destruction reaction rate of $^{16}$O as a function of ion temperature ($T_i$).}
\end{figure}
Fig.~8 reveals that for a spherically symmetric, optically thin plasma with size $R=10^6~\rm{cm}$, and number density $n=10^{18}~\rm{cm^{-3}}$, and for a solar/cosmic initial abundance, the photodisintegration rate is negligible. Even when the electron temperature becomes comparable with the ion temperature, the photodisintegration reaction rate is 3-4 orders of magnitude below other nuclear reaction rates, which is clearly seen by comparing  $^{16}$O$(\gamma, X)$ and $^{16}$O$(p, X)$. In the case when the electron temperature is lower than the ion temperature, and there is a significant amount of neutrons in the plasma (here we took, for simplicity, $n_n=n_p=10^{18}~\rm{cm^{-3}}$), the most significant amount of $\gamma$-rays are produced from $p-n$ interactions. We see from Fig.~8 that the contribution of photodisintegration due to $\gamma$-rays coming from $p-n$ interactions is negligible. We note that nuclear prompt $\gamma$-ray lines have no significant effect in photodisintegration. The same is valid for Deuterium, except if it is in nuclear statistical equilibrium (NSE) as it is in Section.~\ref{subsec:1}. For this case, the equilibrium abundance will strongly depend on electron temperature. If the electron temperature is $T_e<1$~MeV then photodisintegration due to photons emitted by electrons does not play any role, whereas the photodisintegration by photons coming from $p-n$ interactions could affect the NSE equilibrium for timescales longer than $10^4$~s. 

If the size of the plasma and/or electron number density is smaller, photodisintegration will be negligible. However, if the size of the plasma is $R\gtrsim10^8~\rm{cm}$, and/or its number density is $n\gtrsim10^{18}~\rm{cm^{-3}}$, or/and the electron temperature is very high ($T_e>1$~MeV), then photodisntegration must be taken into account.

\subsection{The relevant projectile energy interval}\label{app:ener}

Currently we do not have a complete nuclear theory which allows to calculate precisely the $\sigma_{ij}^k$ for any requested energy interval. Therefore, a good approximation must be used here.

We are interested in plasma reactions in the temperature interval $0.1-20~\rm{MeV}$. For any given temperature, there exists an energy interval or window, such that, integration over it, brings the major contribution to the Maxwellian average (for that particular temperature).
 
$$\int_0^\infty\,=\,\int_0^{\epsilon_1(T)}\,+\,\left[\int_{\epsilon_1(T)}^{\epsilon_2(T)}\right]\,+\,\int_{\epsilon_2(T)}^\infty$$
$$\frac{\int_0^{\epsilon_1(T)}\,+\,\int_{\epsilon_2(T)}^\infty}{\int_{\epsilon_1(T)}^{\epsilon_2(T)}}\equiv \varepsilon\sim0.1\%$$
Working within the temperature range of $0.1-20$~MeV, we want to obtain the range of the integration $[E_1,E_2]$ such that the accuracy of the Maxwellian average is greater than a desired value ($\epsilon$). The energy interval $[E_1,E_2]$, will contain any energy window $[\epsilon_1(T),\epsilon_2(T)]$ for any temperatures in $0.1-20$~MeV. It is expected that the lowest temperature ($T_1$) will affect the value the lowest energy $E_1$, and the highest temperature ($T_2$) that of the energy $E_2$. This is true due to the distribution functions of these two extreme temperatures.
 
Bearing in mind the above argument, we can estimate $E_1$ and $E_2$ from $T_1$ and $T_2$, separately. For simplicity, let us consider an extreme case of constant cross section with a value equal to the peak of the actual cross section. Since the energy  window upper limit ($\epsilon_2$) for $T_1$ and the energy window lower limit ($\epsilon_1$) for $T_2$ lies inside $[E_1,E_2]$ we can make the following approximations to the integral in Eq.(\ref{rf}):

$$\frac{\int_0^{E_1}\,dE\,E\,\exp(-E/T_1)}{\int_{E_1}^\infty\,dE\,E\,\exp(-E/T_1)}\equiv \varepsilon$$
$$\frac{\int_{E_2}^\infty\,dE\,E\,\exp(-E/T_2)}{\int_0^{E_2}\,dE\,E\,\exp(-E/T_2)}\equiv \varepsilon$$

Integrating the above equations we obtain the following algebraical equations that allow to find the root:

$$(1+\frac{E_1}{T_1})\,\exp(-\frac{E_1}{T_1})=\frac{1}{1+\varepsilon}$$
$$(1+\frac{E_2}{T_2})\,\exp(-\frac{E_2}{T_2})=\frac{\varepsilon}{1+\varepsilon}.$$
If we take $T1=0.1$~MeV, $T_2=20$~MeV, and $\varepsilon=0.1\%$, the solutions are
$$E_1\approx 0.015\,\text{MeV} \qquad\qquad E_2\approx 133\,\rm{MeV}$$

For less accurate results, the interval $[E_1,E_2]$ will be narrowed. For the capture reactions such as $(n,\gamma)$, where the cross sections at low energies show a dependence $\sigma\sim1/v$, it will be a more complicated function but the lower valid energy holds for $\varepsilon=1\%$.

The last estimation informs us numerically of the energy  interval beyond which the contribution is negligible. This interval is $E\in\,[0.01,\,150]$ MeV for accuracy $\leq1\%$. The better we know the cross sections in this interval, the more accurate the values of the reaction rates will be. 
%The only remaining issue is to calculate the reaction cross sections in this interval of energy.

\section{Cross sections and reaction rates for some typical nuclei}\label{app:plot}
\subsection{Total nonelastic and inelastic cross sections with their corresponding rates}
% \centering
\includegraphics[scale=0.47]{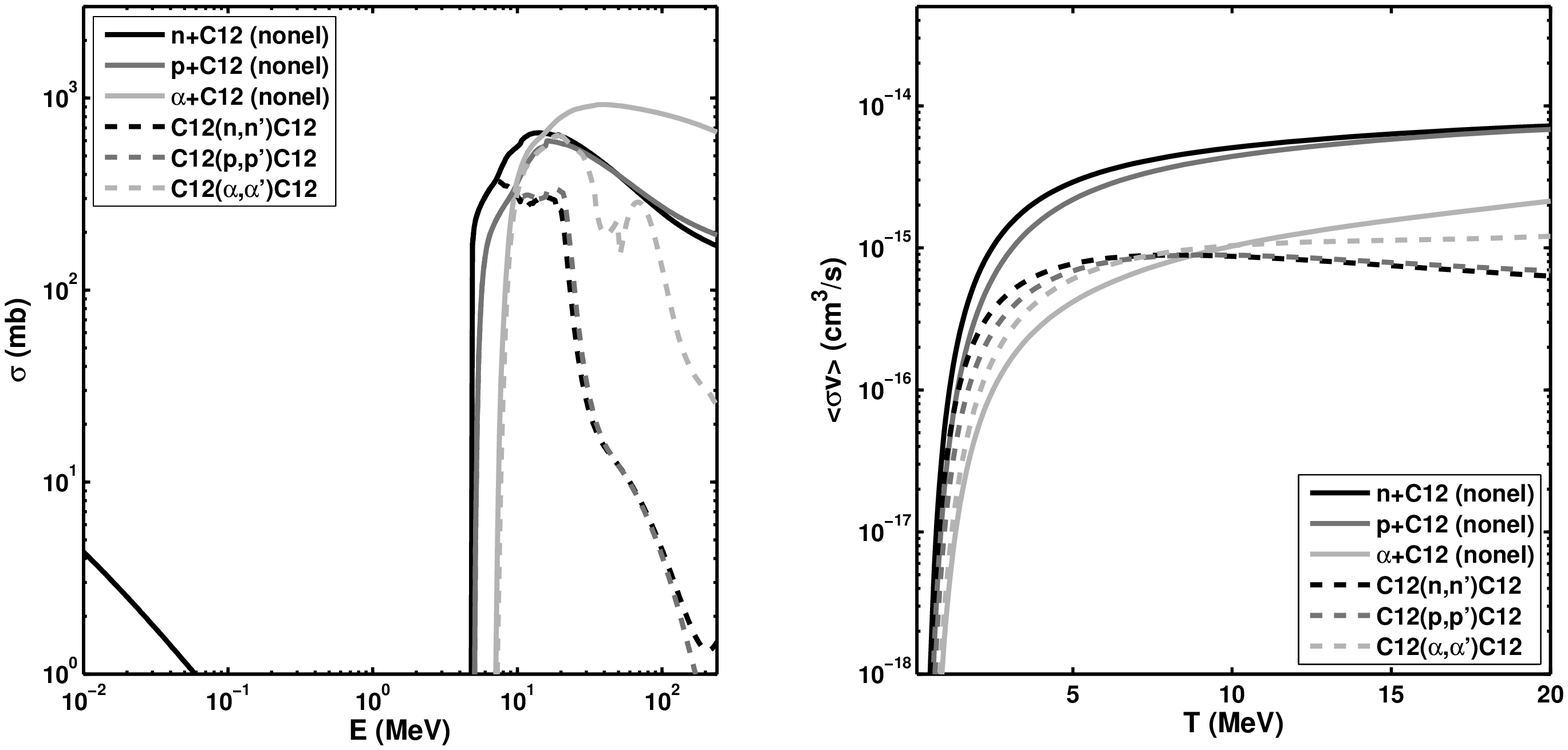}
\includegraphics[scale=0.47]{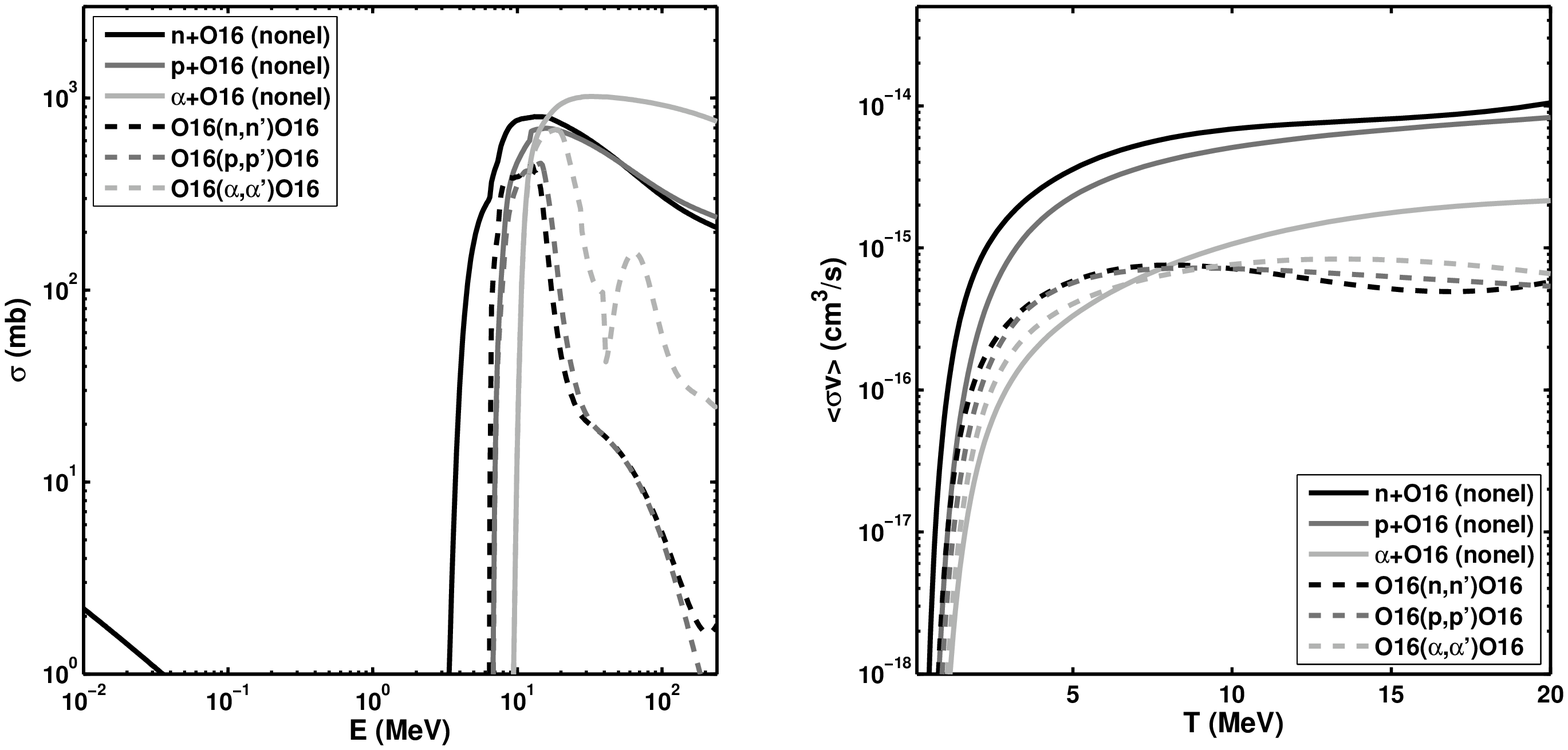}
\includegraphics[scale=0.47]{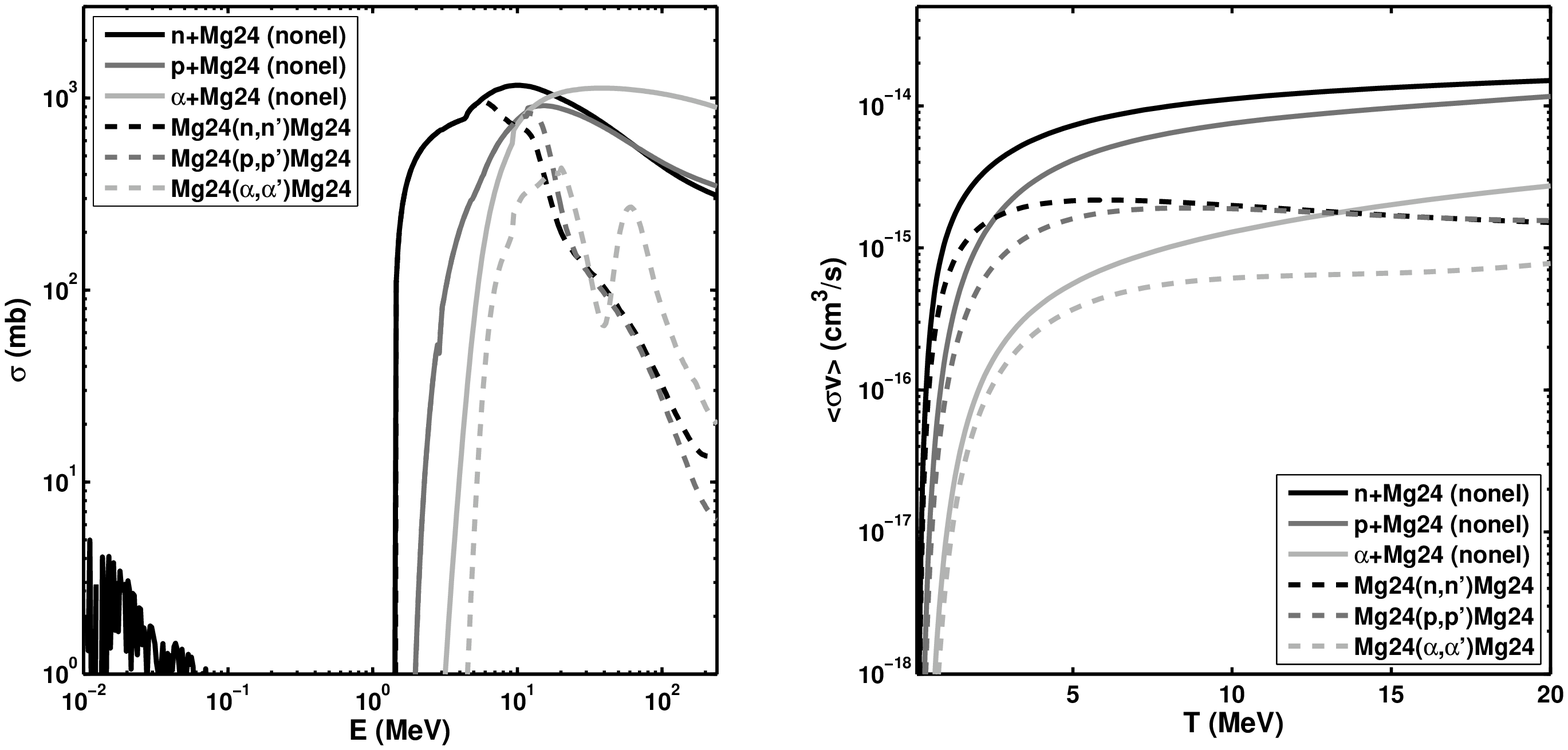}
\includegraphics[scale=0.47]{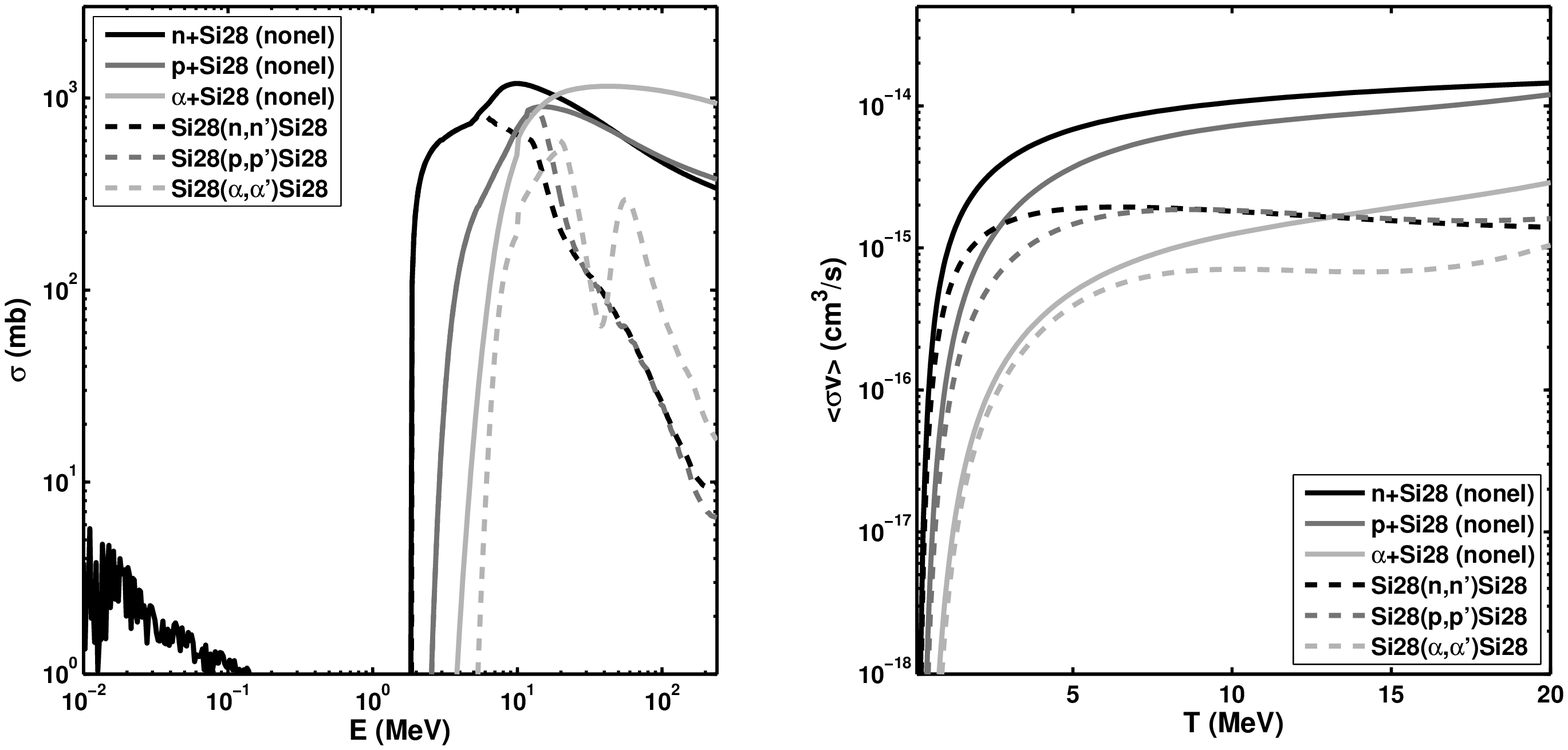}
\includegraphics[scale=0.47]{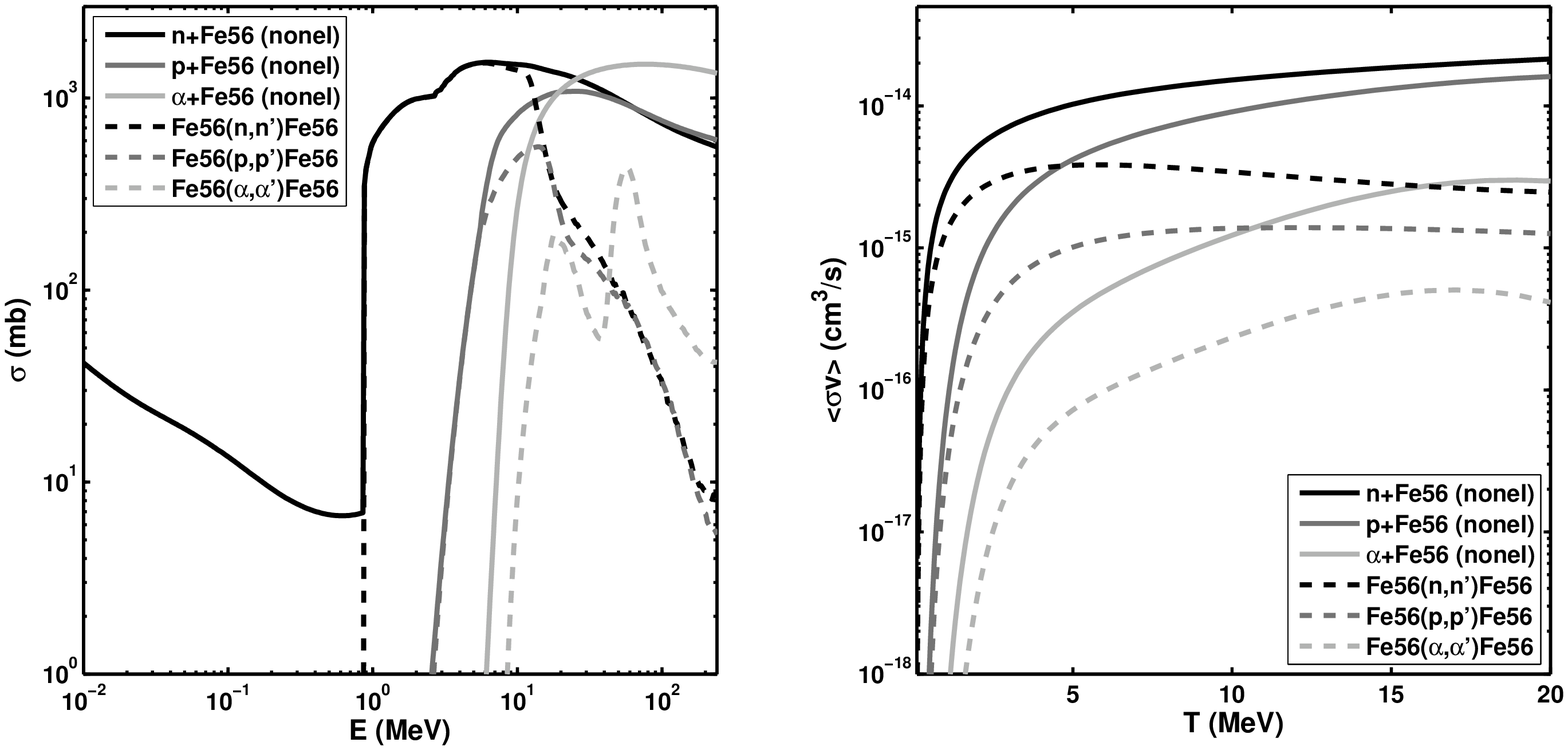}

\subsection{Comparison between inelastic and capture cross sections for some specific elements}
% \centering
\includegraphics[scale=0.5]{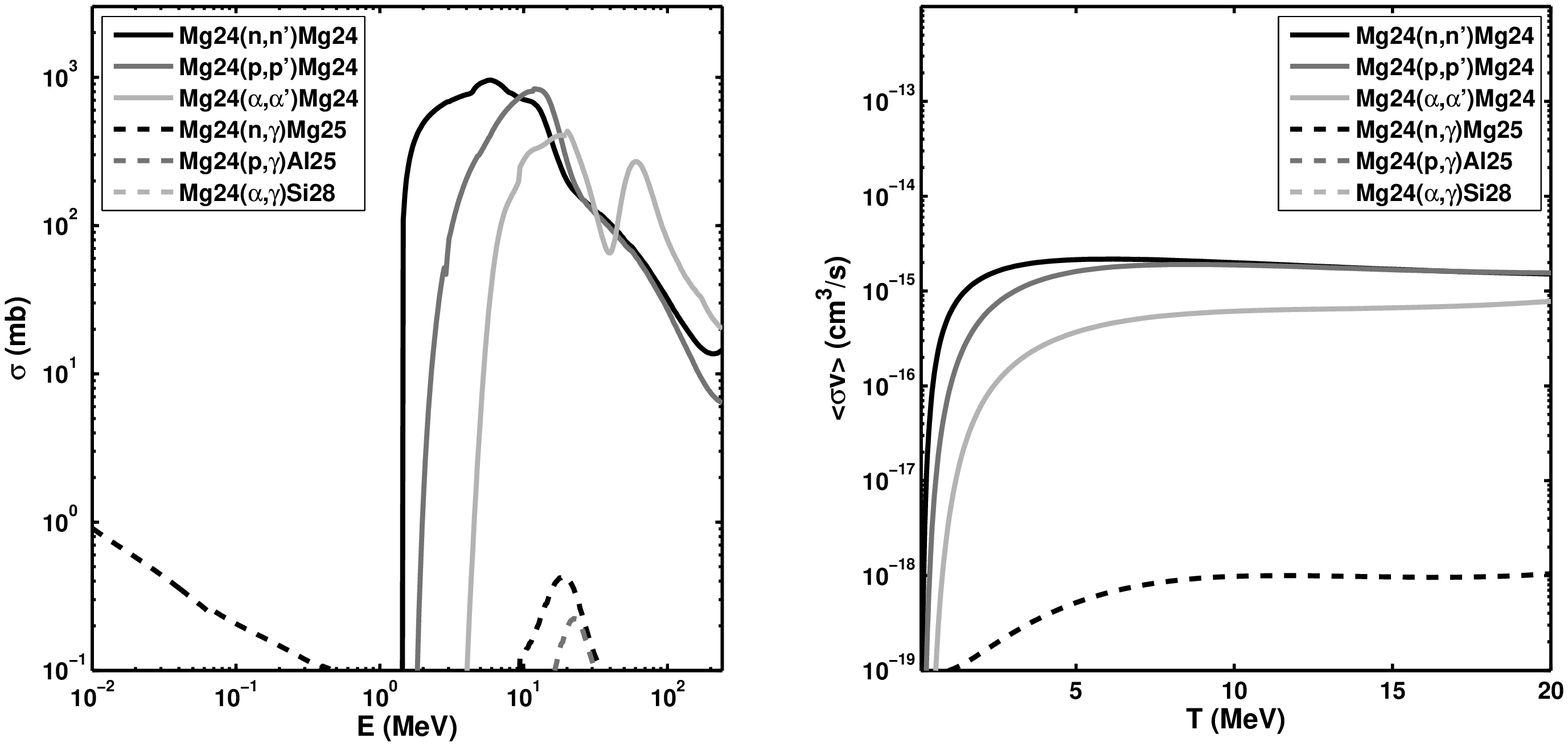}
\includegraphics[scale=0.5]{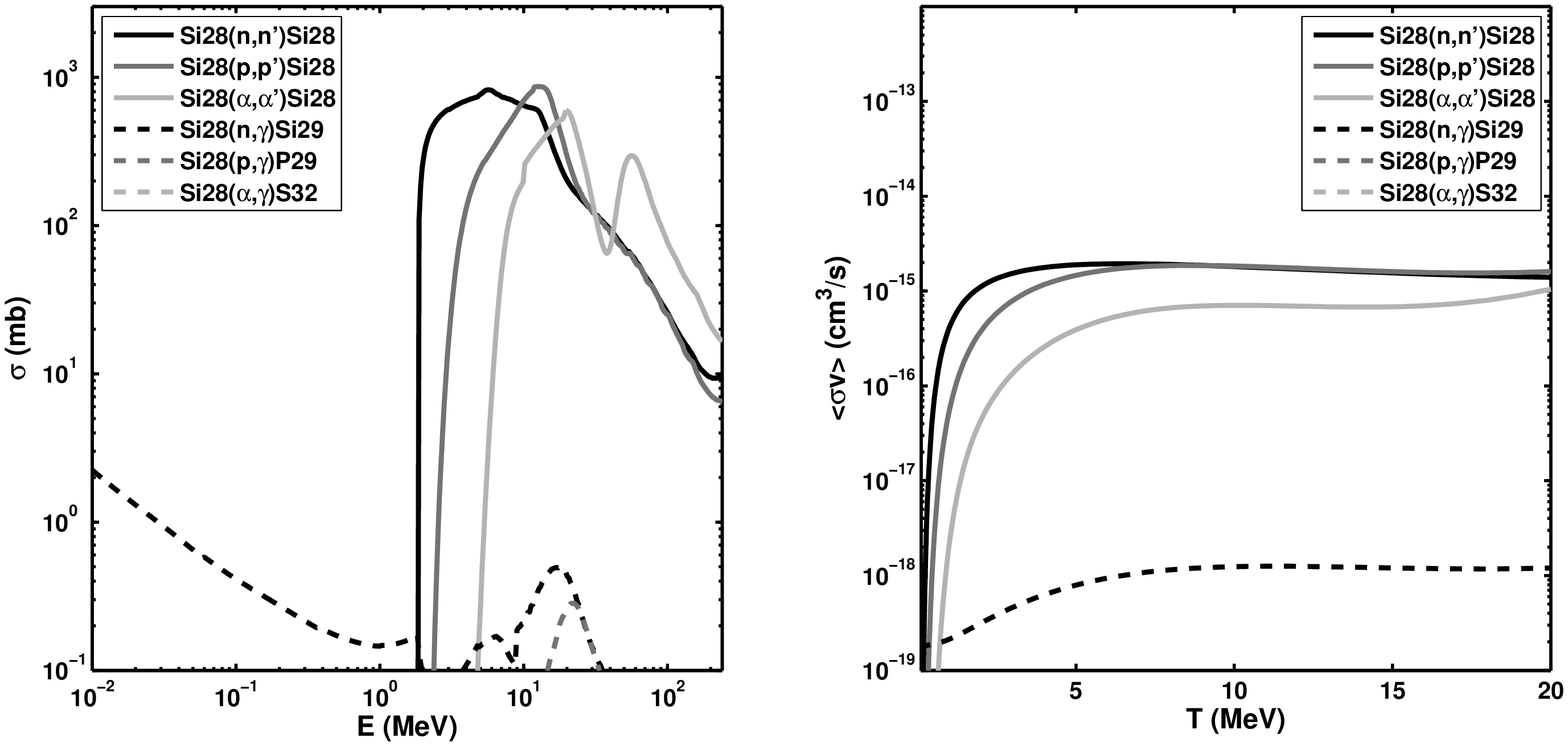}
\includegraphics[scale=0.5]{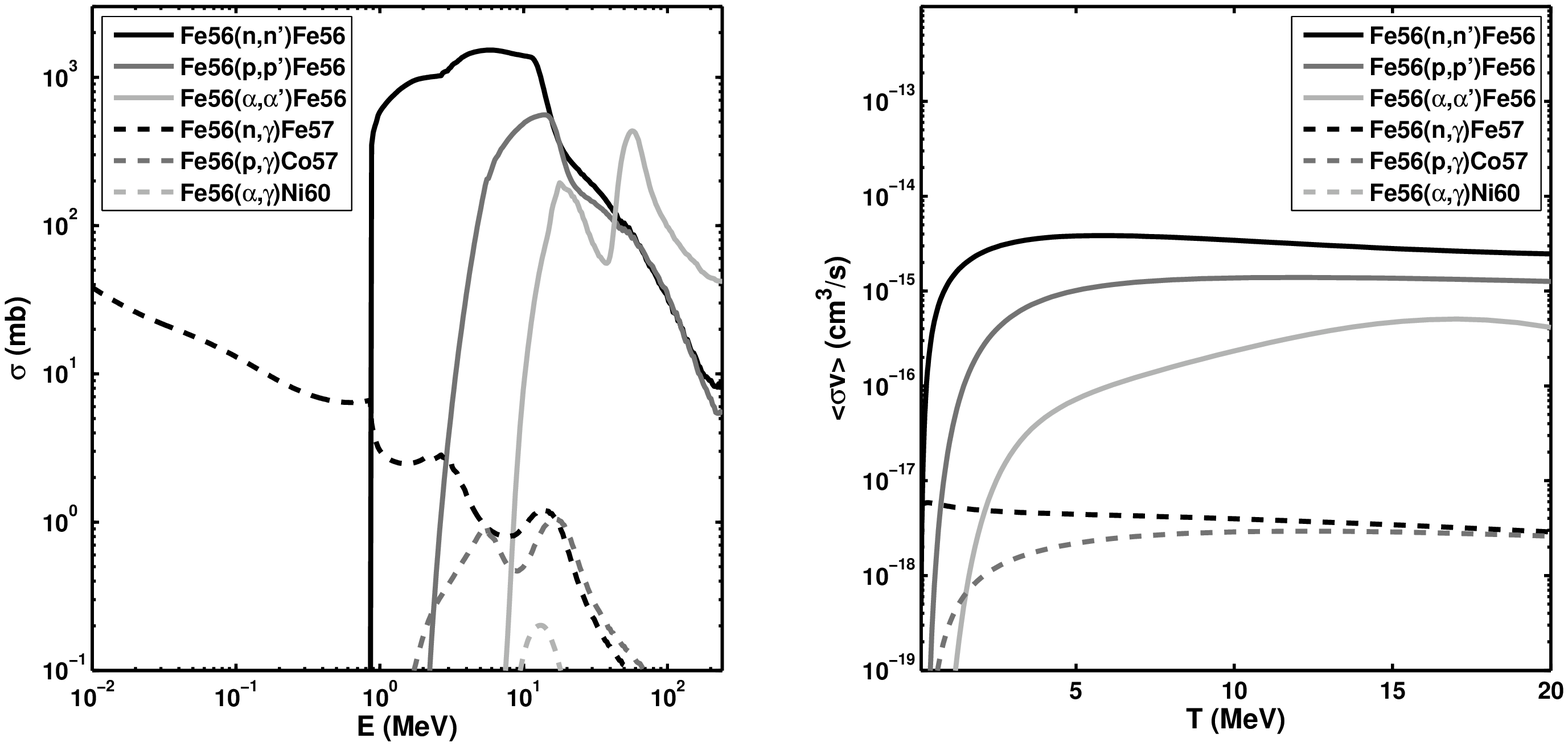}

\subsection{Production of $\gamma$-ray lines for some elements}
\centering
\includegraphics[scale=0.8]{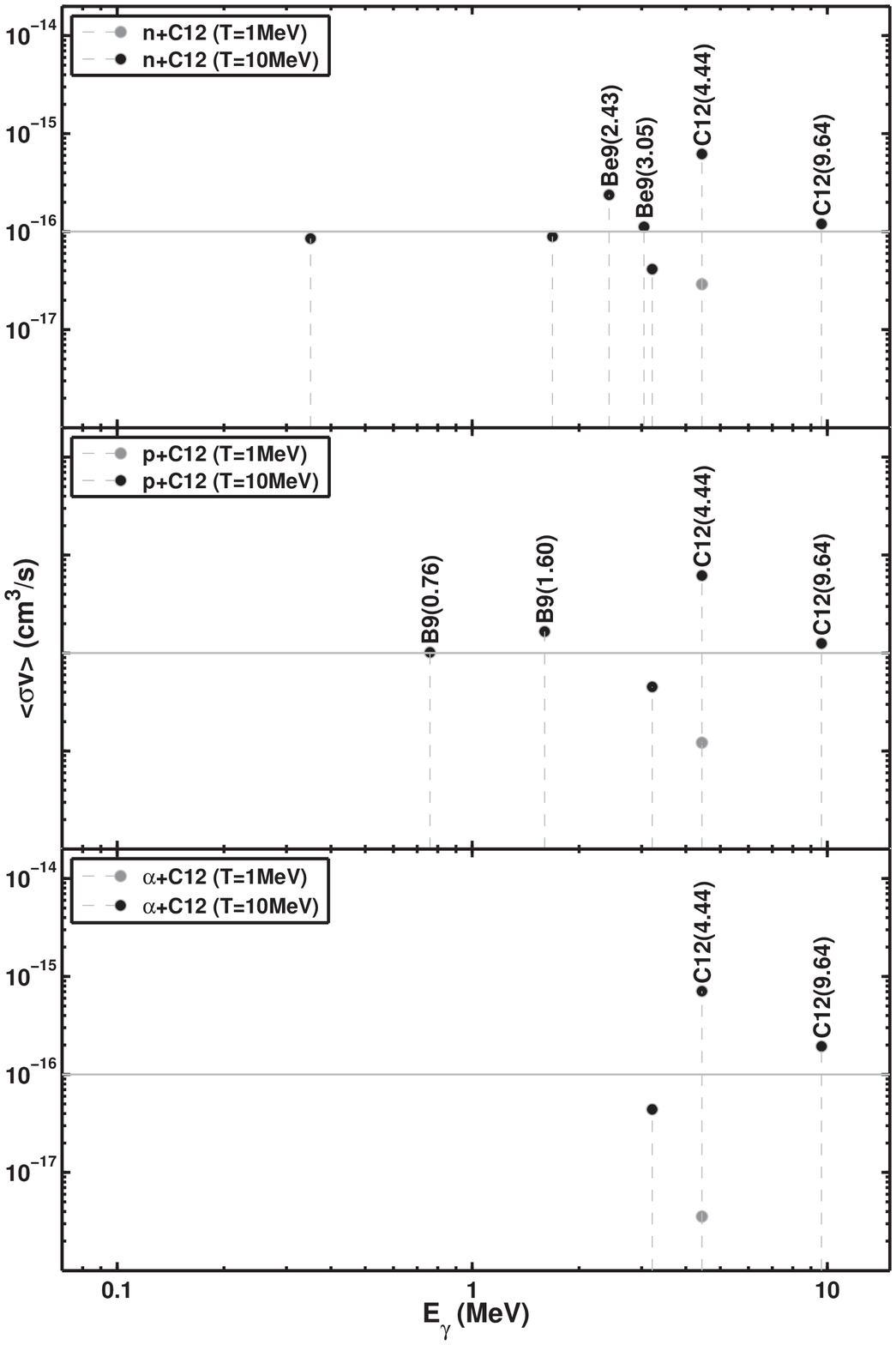}\\
\includegraphics[scale=0.52]{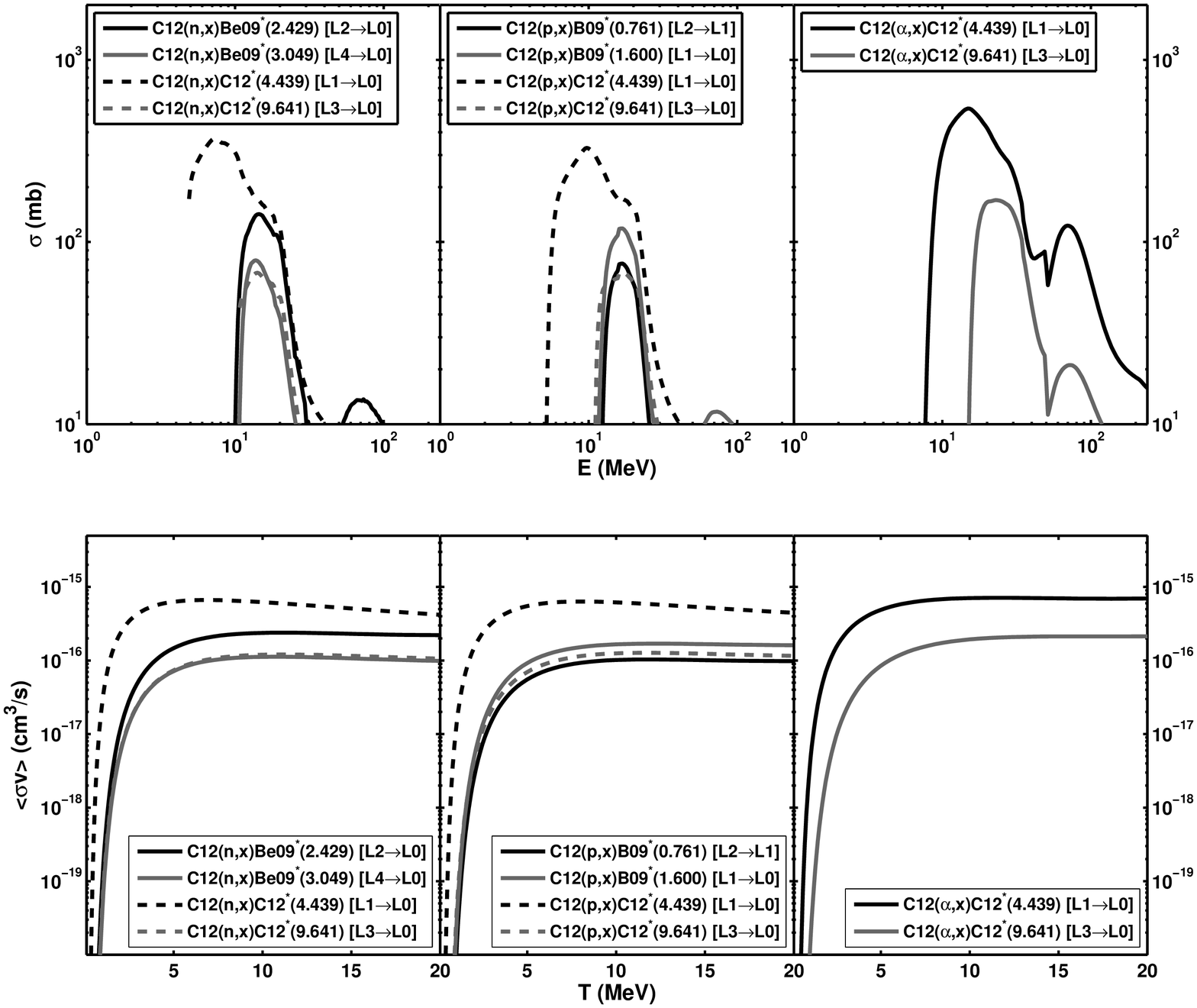}\\
\includegraphics[scale=0.8]{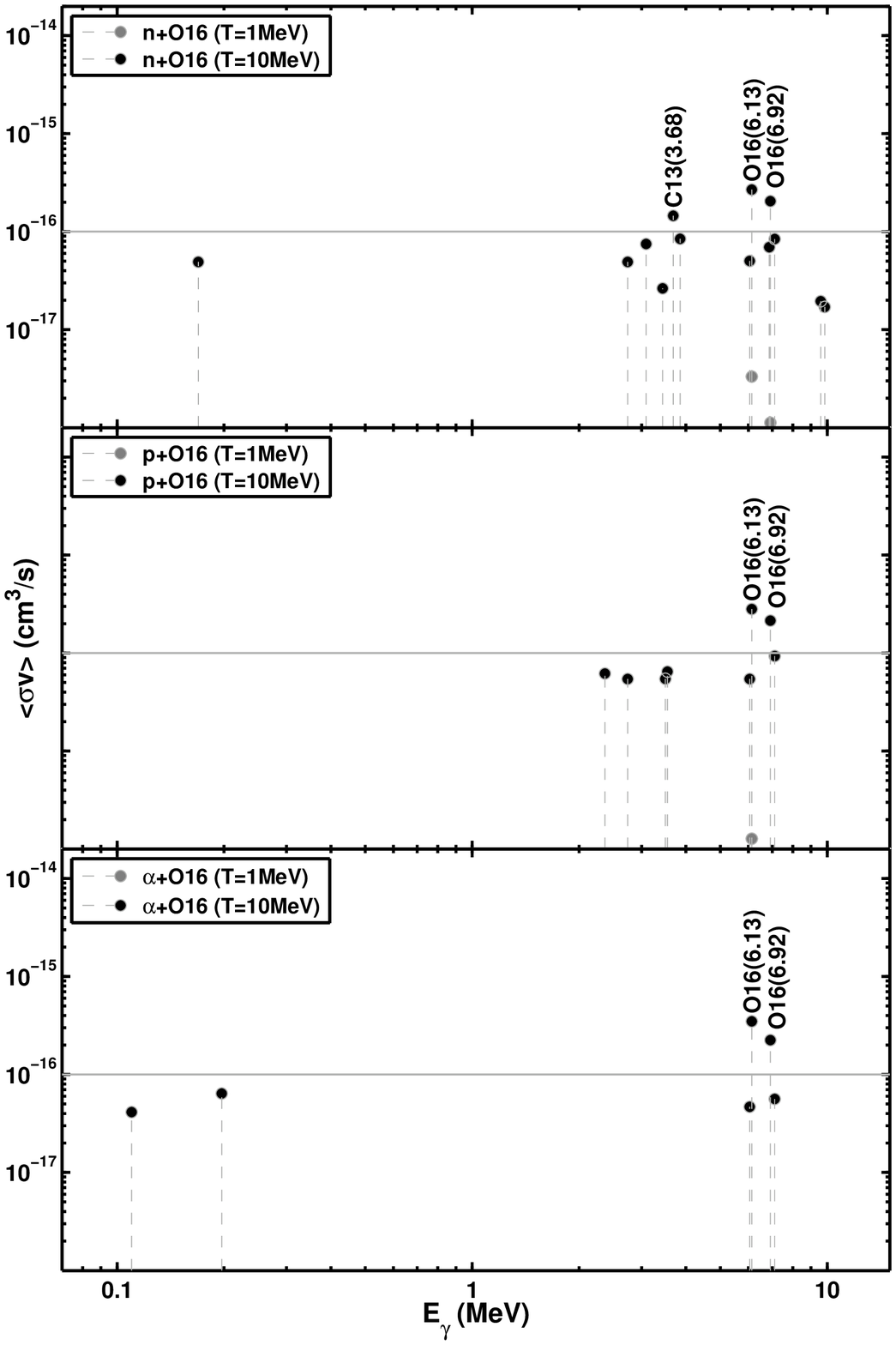}\\
\includegraphics[scale=0.52]{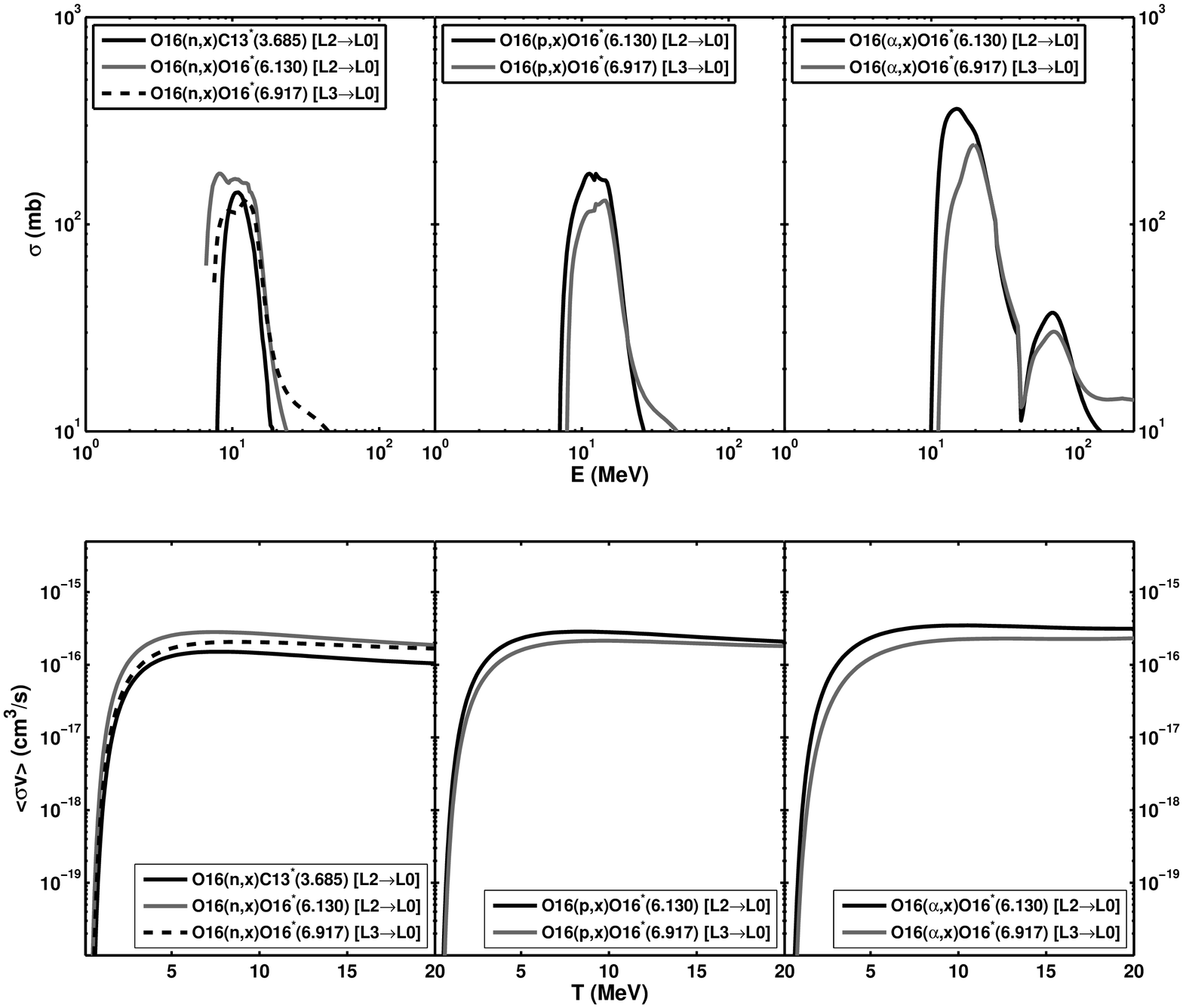}\\
\includegraphics[scale=0.8]{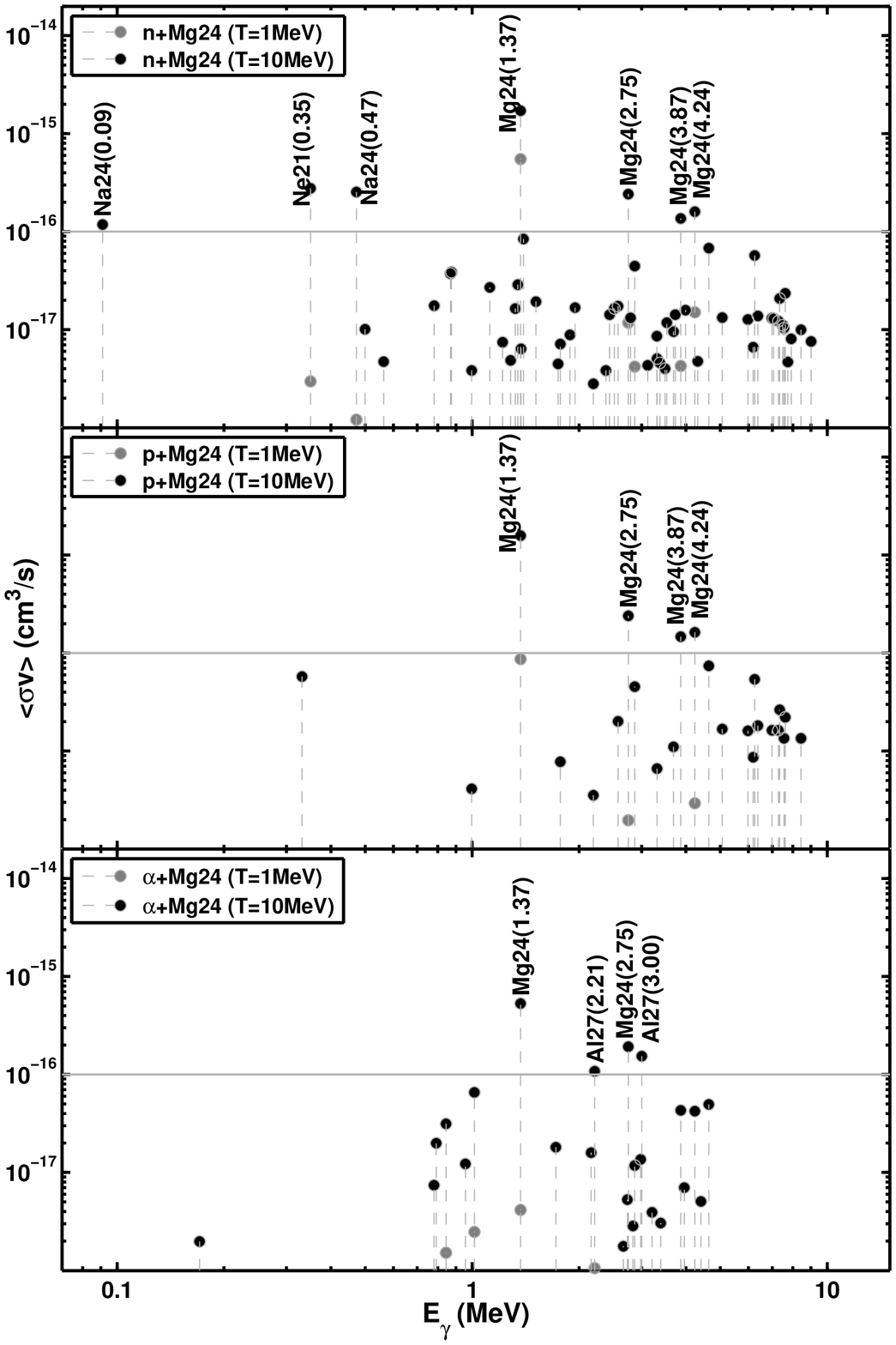}\\
\includegraphics[scale=0.52]{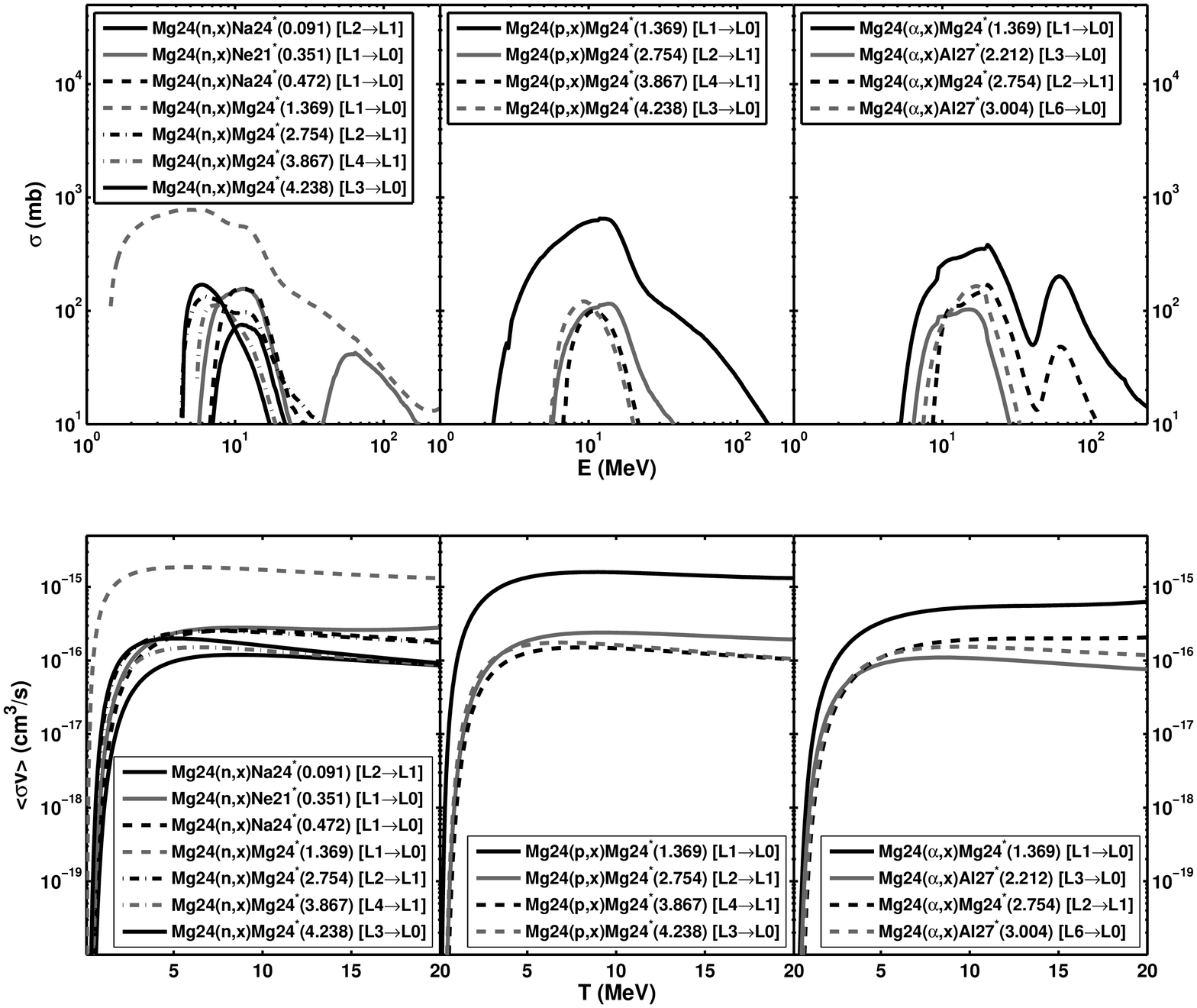}\\
\includegraphics[scale=0.8]{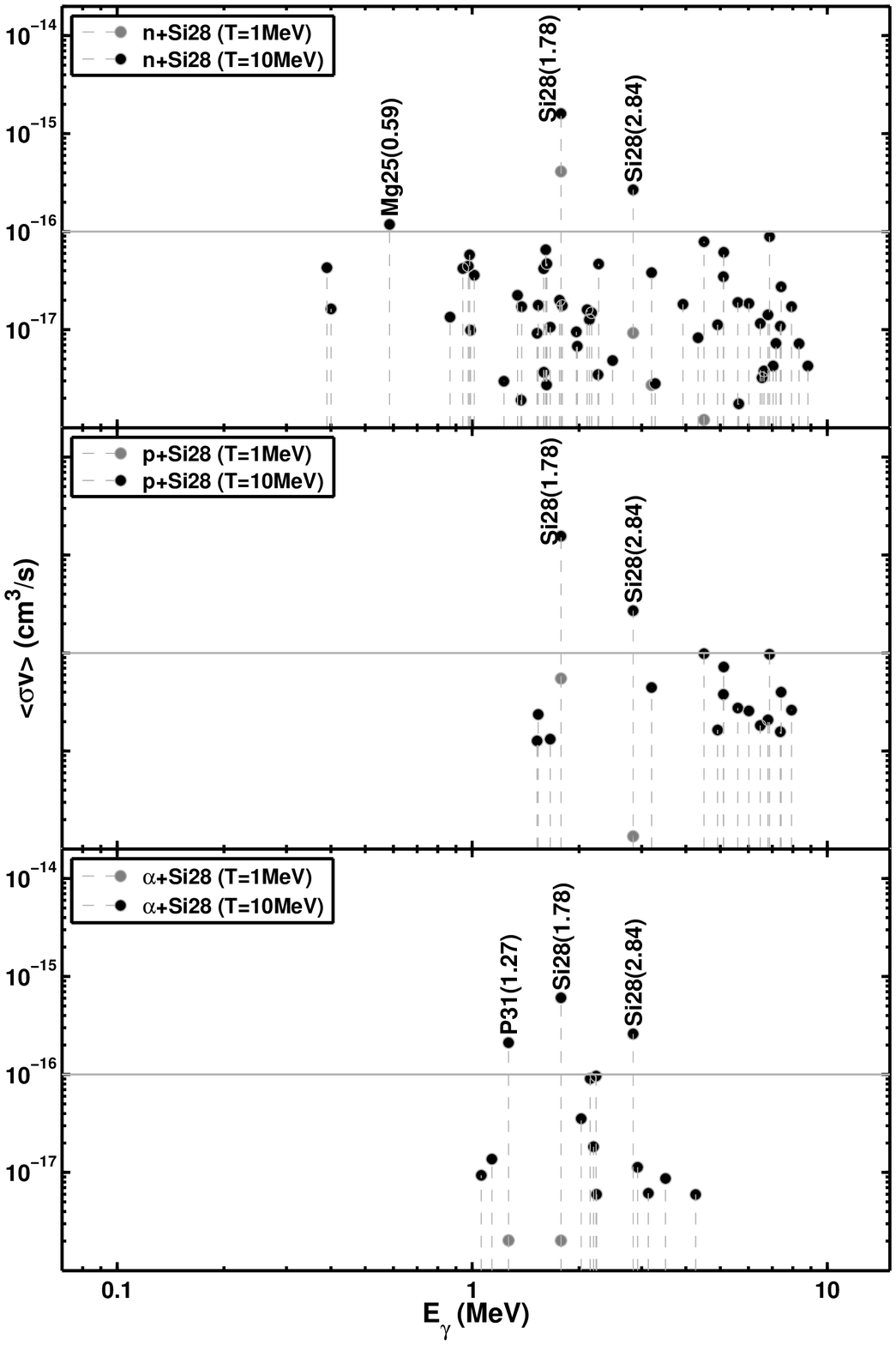}\\
\includegraphics[scale=0.52]{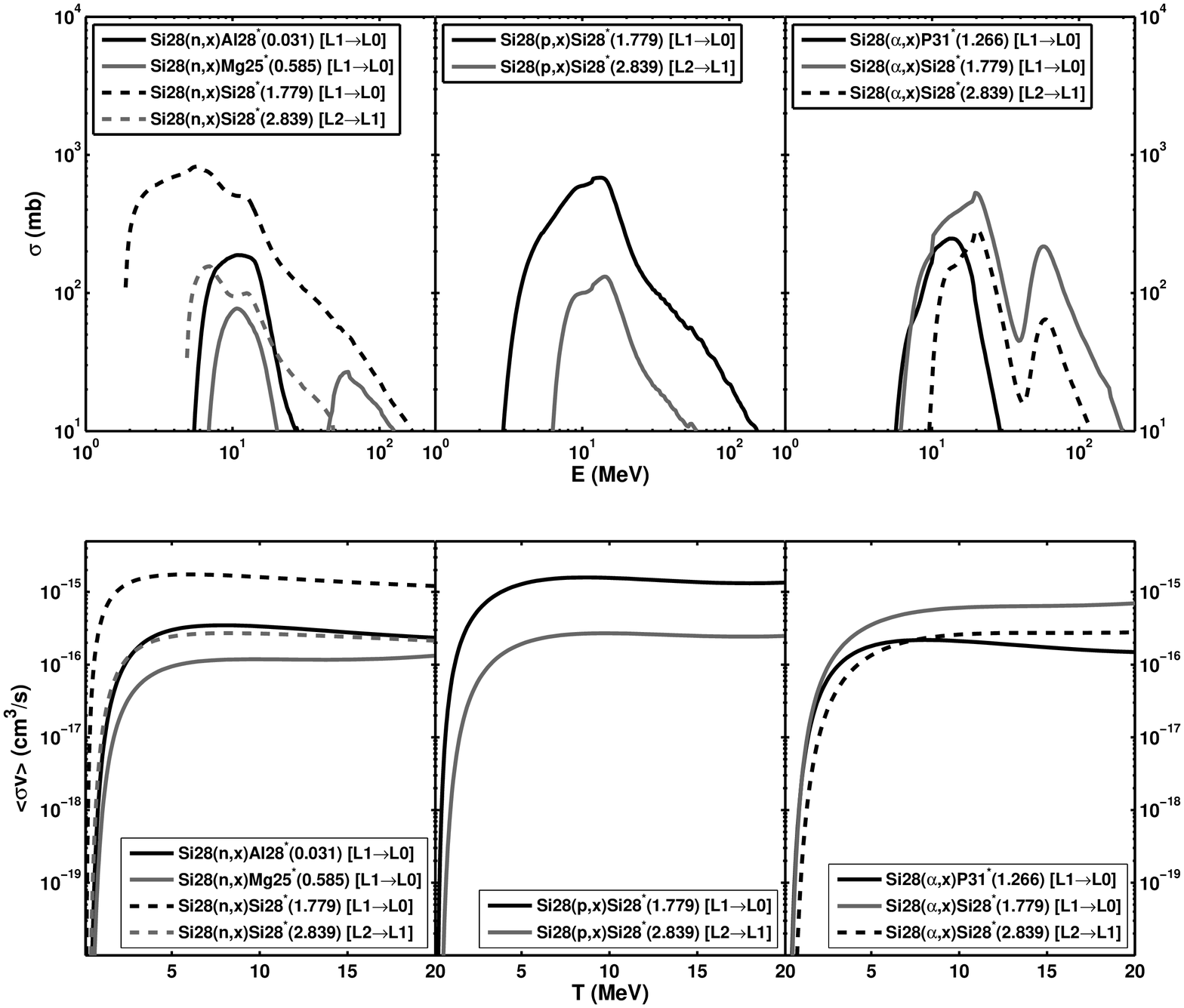}\\
\includegraphics[scale=0.8]{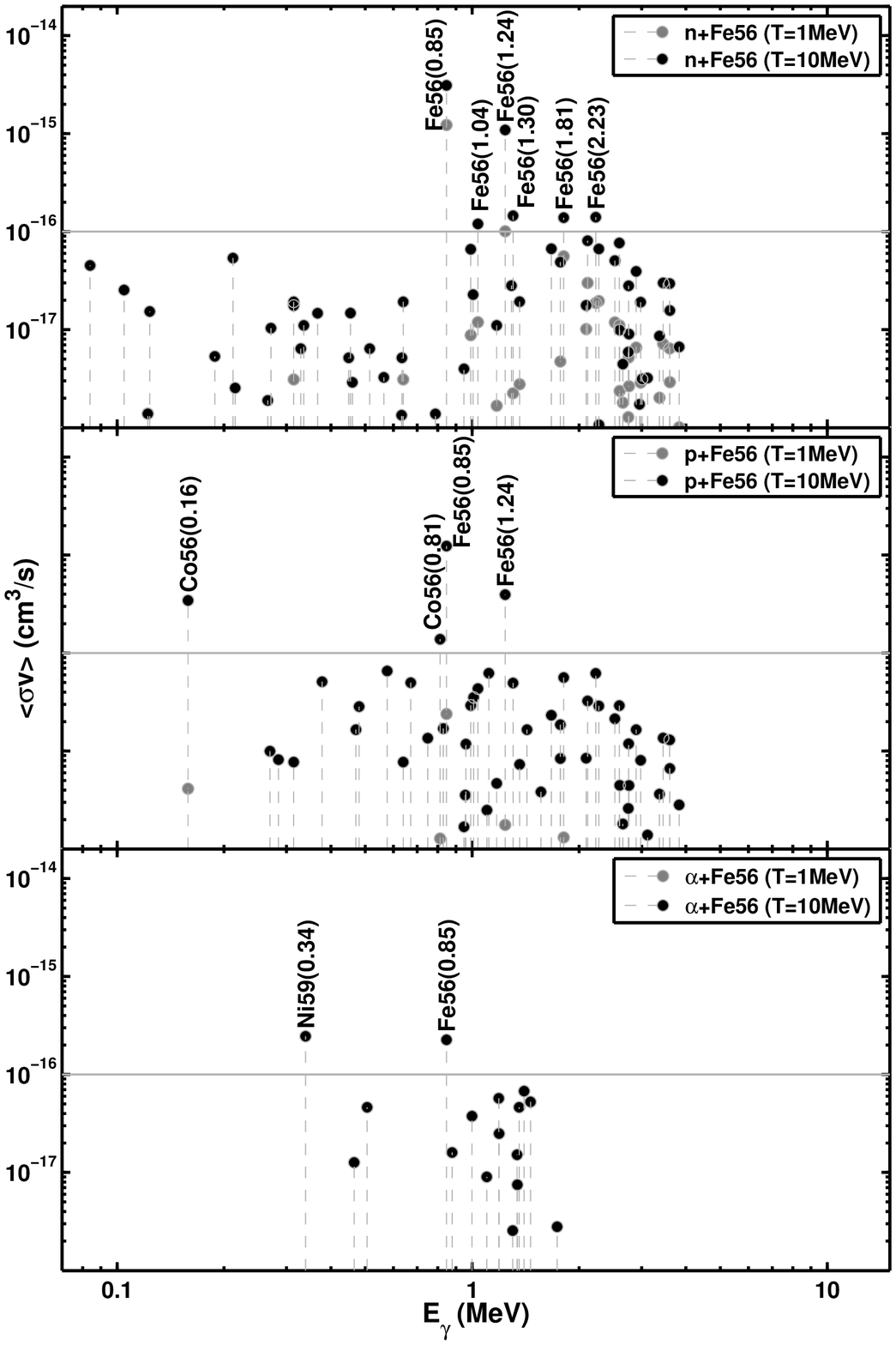}\\
\includegraphics[scale=0.52]{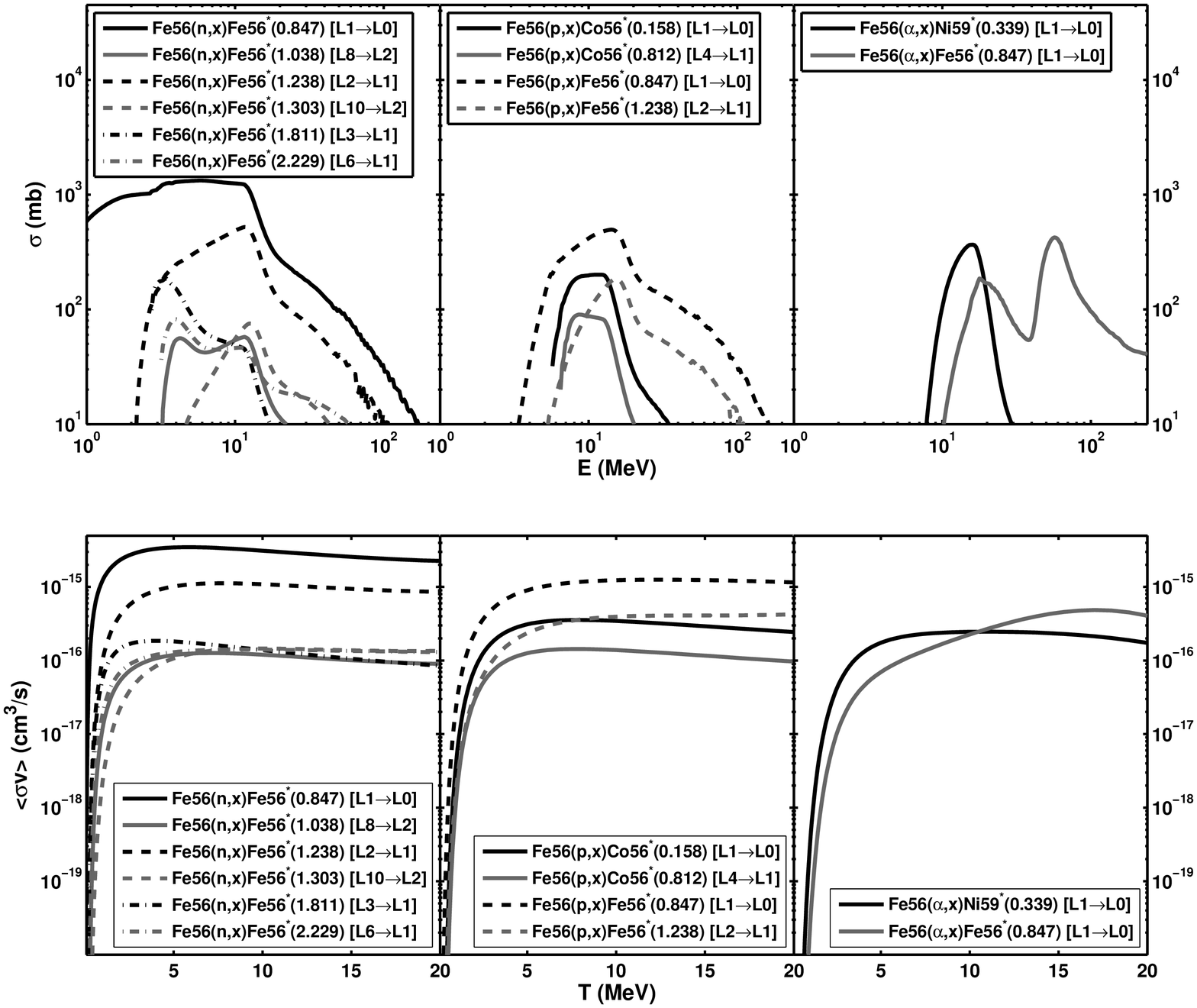}\\

%\begin{thebibliography}{000} %for 3 digits

\end{document}